\documentclass[twocolumn,aps,prb,superscriptaddress]{revtex4}
\usepackage{amsmath}
\usepackage{amssymb}
\usepackage{graphicx}
\usepackage{color}

\newcommand{\bea}{\begin{eqnarray}}
\newcommand{\eea}{\end{eqnarray}}
\newcommand{\bse}{\begin{subequations}}
\newcommand{\ese}{\end{subequations}}

\newcommand{\tcs}{${\rm ThCr_2Si_2}$}
\newcommand{\cca}{CaCo$_{2-y}$As$_2$}
\newcommand{\sca}{${\rm SrCo_2As_2}$}
\newcommand{\bfa}{${\rm BaFe_2As_2}$}
\newcommand{\bca}{${\rm BaCo_2As_2}$}

\newcommand{\csca}{Ca$_{1-x}$Sr$_x$Co$_{2-y}$As$_2$}

\newcommand{\scna}{Sr(Co$_{1-x}$Ni$_x$)$_2$As$_2$}
\newcommand{\ccia}{Ca(Co$_{1-x}$Ir$_x$)$_{2-y}$As$_2$}

\begin{document}

\title{Ferromagnetic Cluster-Glass Phase in Ca(Co$_{1-x}$Ir$_x$)$_{2-y}$As$_2$ crystals}

\author{Santanu Pakhira}
\affiliation{Ames Laboratory, Ames, Iowa 50011, USA}
\author{N. S. Sangeetha}
\affiliation{Ames Laboratory, Ames, Iowa 50011, USA}
\author{V. Smetana}
\affiliation{Department of Materials and Environmental Chemistry, Stockholm University, Svante Arrhenius v\"{a}g 16 C, 106 91 Stockholm, Sweden}
\author{A.-V. Mudring}
\affiliation{Department of Materials and Environmental Chemistry, Stockholm University, Svante Arrhenius v\"{a}g 16 C, 106 91 Stockholm, Sweden}
\author{D. C. Johnston}
\affiliation{Ames Laboratory, Ames, Iowa 50011, USA}
\affiliation{Department of Physics and Astronomy, Iowa State University, Ames, Iowa 50011, USA}

\date{\today}

\begin{abstract}
Single crystals of \ccia\ with $0\leq x \leq 0.35$ and $0.10\leq y \leq 0.14$ have been grown using the self-flux technique and  characterized by single-crystal x-ray diffraction (XRD), energy-dispersive x-ray spectroscopy, magnetization $M$ and magnetic susceptibility $\chi$ measurements versus temperature~$T$, magnetic field~$H$, and time~$t$, and heat capacity $C_{\rm p}(H,T)$ measurements. The XRD refinements reveal that all the Ir-substituted crystals crystallize in a collapsed-tetragonal structure as does the parent \cca\ compound. A small 3.3\% Ir substitution for Co in CaCo$_{1.86}$As$_2$ drastically lowers the A-type antiferromagnetic (AFM) transition temperature~$T_{\rm N}$ from 52 to 23~K with a significant enhancement of the Sommerfeld electronic heat-capacity coefficient. The A-type AFM structure consists of $ab$-plane layers of spins ferromagnetically-aligned along the $c$~axis with AFM alignment of the spins in adjacent layers along this axis.  The positive Weiss temperatures obtained from Curie-Weiss fits to the $\chi(T>T_{\rm N})$ data indicate that the dominant magnetic interactions are ferromagnetic (FM) for all~$x$.  A magnetic phase boundary is inferred to be present between $x = 0.14$ and $x=0.17$ from a discontinuity in the $x$ dependences of the effective moment and Weiss temperature in the Curie-Weiss fits.  FM fluctuations that strongly increase with increasing~$x$ are also  revealed from the $\chi(T)$ data. The magnetic ground state for $x \geq 0.17$ is a spin glass as indicated by hysteresis in $\chi(T)$ between field-cooling and zero-field-cooling measurements and from the relaxation of $M$ in a small field that exhibits a stretched-exponential time dependence.  The spin glass has a small FM component to the ordering and is hence inferred to be comprised of small FM clusters. The competing AFM and FM interactions along with crystallographic disorder associated with Ir substitution are inferred to be responsible for the development of the FM cluster-glass phase. A logarithmic $T$ dependence of $C_{\rm p}$ at low~$T$ for $x = 0.14$ is consistent with the presence of significant FM quantum fluctuations. This composition is near the $T = 0$ boundary at $x\approx 0.16$ between the A-type AFM phase containing ferromagnetically-aligned layers of spins and the FM cluster-glass phase.

\end{abstract}

\maketitle

\section{Introduction}

Since the discovery of high-$T_{\rm c}$ superconductivity (SC) in doped iron arsenides $A$Fe$_2$As$_2$ ($A$ = Ca, Sr, Ba, Eu), the interplay between magnetism and SC in these materials opened up new research areas~\cite{Rotter2008, Johnston2010, Canfield2010, Paglione2010, Fernandes2010, Stewart2011, Scalapino2012, Dai2012, Dagotto2013, Dai2015}. Ternary arsenide compounds having the general formula $AM_2$As$_2$ ($M$ = transition metal) commonly crystallize in the layered body-centered-tetragonal ThCr$_2$Si$_2$-type crystal structure. Here $M_2$As$_2$ layers are comprised of edge-sharing $M$As$_4$ tetrahedra and the $A$ atoms occupy layers between the $M_2$As$_2$ layers. Depending on the ratio $c/a$ of the tetragonal $c$ and~$a$ lattice parameters, a compound can crystallize in either the collapsed-tetragonal (cT) or uncollapsed-tetragonal (ucT) version of the structure or in the intermediate regime~\cite{Anand2012}. Electron and/or hole doping in the three different atomic sites have been reported to introduce superconducting and/or magnetic phenomena in these materials. The Fe-based arsenide compounds have been extensively studied after the discovery of SC at $T_c$ = 38 K in K-doped \bfa\ in 2008~\cite{Rotter2008}. Both hole doping and electron doping subsequently resulted in the observation of SC in the $A$Fe$_2$As$_2$ ($A$ = Sr, Ca, and Eu)  systems~\cite{Sasmal2008, Jasper2008, Kumar2009, Ren2009}. In these systems superconductivity is realized by suppressing the long-range antiferromagnetic (AFM) order of the parent compounds through chemical substitution or by the application of pressure. Later it was found that the suppression of long range AFM order while preserving strong dynamic short-range AFM correlations was required for SC to appear, indicating that AFM fluctuations are required for the appearance of SC in the iron arsenides~\cite{Johnston2010, Canfield2010, Paglione2010, Stewart2011, Scalapino2012, Dai2012, Dagotto2013,Dai2015}. The self-doped alkali-metal compounds  KFe$_2$As$_2$ ($T_{\rm c} = 3.8$~K), RbFe$_2$As$_2$ ($T_{\rm c} = 2.6$~K), and CsFe$_2$As$_2$ ($T_{\rm c} = 2.6$~K) also exhibit SC~\cite{Sasmal2008, Rotter2008Angew, Bukowski2010}. These discoveries sparked interest to study other transition-element-based analogues of this family of materials.

Metallic \tcs-type CoAs-based compounds exist but do not exhibit SC\@.  However, these materials have attracted significant interest due to their peculiar itinerant magnetic behaviour arising from their electronic structure and sensitivity to chemical substitution.  For example, metallic \bca\ has an ucT structure and does not exhibit long-range magnetic ordering down to a temperature $T=1.8$~K~\cite{Sefat2009,Anand2014}. However, crystals of this compound have a rather large magnetic susceptibility $\chi$ with a broad maximum followed by a weak low-temperature upturn~\cite{Sefat2009}. A large exchange-enhanced density of states at the Fermi energy ${\cal D}(E_{\rm F}) \approx 18$~states/eV~f.u.\ was estimated for the material from low-$T$ heat capacity data, where f.u.\ stands for formula unit~\cite{Anand2014}. In the initial report, it was argued that long-range ferromagnetic (FM) ordering is suppressed by quantum fluctuations~\cite{Sefat2009}, although the subsequent study~\cite{Anand2014} showed that the properties of \bca\ are not very sensitive to chemical doping.

On the other hand, both \sca\ and \cca\ crystallize in the cT structure~\cite{Jasper2008, Anand2012, Pandey2013, Quirinale2013, Anand2014Ca}. Metallic  \sca\ is a Stoner-enhanced paramagnet with no long-range magnetic ordering or SC at temperatures above 0.05~K~\cite{Bing2019sca}. Inelastic neutron-scattering measurements revealed the presence of stripe-type AFM fluctuations at the same wave vector as observed for the high-$T_{\rm c}$ parent compounds $A$Fe$_2$As$_2$~\cite{Jayasekara2013, Bing2019sca}. However, in contrast to the $A$Fe$_2$As$_2$ compounds, no obvious Fermi-surface nesting was observed at that wave vector for \sca\ and furthermore, strong FM fluctuations occur that evidently hinder the occurrence of SC~\cite{Wiecki2015, Li2019}.  Such FM fluctuations were also observed in FeAs-based superconductors, suggesting that the different $T_{\rm c}$'s observed in these materials may be at least partially explained by different levels of FM fluctuations in the compounds~\cite{Wiecki2015b}.

In contrast to \bca\ and \sca, a detectable concentration of vacancies is observed on the Co sites in \cca\ that undergoes A-type AFM ordering at $T_{\rm N} = 52$--77~K, depending on the crystal~\cite{Anand2014Ca, Cheng2012, Ying2012}. In this magnetic structure, the ordered moments on the Co sites within the $ab$ plane are ferromagnetically aligned along the $c$~axis with AFM alignment between adjacent planes. FM interactions dominate together with relatively weak AFM interactions between Co planes responsible for the A-type AFM ordering. Inelastic neutron scattering experiments reveal the presence of strong magnetic frustration in the system within the $J_1$-$J_2$ Heisenberg model on a square lattice with the  nearest-neighbour exchange interaction being ferromagnetic type~\cite{Sapkota2017}. This strong magnetic frustration establishes \cca\ as a unique member of the ternary arsenide family.

Chemically-doped Co-based arsenides are of significant interest due to the interplay between the lattice, electronic and magnetic degrees of freedom. For example, the K-doped compound Ba$_{0.94}$K$_{0.06}$Co$_2$As$_2$ shows weak FM behaviour; however, the magnetic behavior of the  composition Ba$_{0.78}$K$_{0.22}$Co$_2$As$_2$ is found to be quite similar to that observed for undoped \bca~\cite{Anand2014}. This difference has been suggested to be due to different magnetic defect levels associated with the crystal growth. It has also been discovered that the system \csca\ exhibits a composition-induced crossover in the magnetic anisotropy~\cite{Ying2013,Sangeetha2017}. In the region $0 \leq x \leq 0.2$, the compounds order in the A-type AFM structure (AFM I phase) where the moments are aligned along the $c$-axis. The compounds with $x$ = 0.40 and 0.45 order in the AFM II structure with the ordered moments aligned in the $ab$-plane. The composition with $x = 0.33$ exhibits a temperature-induced transition between moment alignments in the $ab$~plane and along the $c$~axis. Neutron diffraction measurements yield the presence of magnetic frustration within the Co layers in \csca~\cite{Bing2019csca}.

Recently, non-Fermi-liquid types of behavior associated with a composition-induced magnetic quantum-critical point in \scna\ crystals near $x = 0.3$ has been reported~\cite{Sangeetha2019scna}. In addition, crystals of \scna\ with $0 < x < 0.3$ exhibit $c$-axis helical AFM ordering~\cite{Sangeetha2019scna, YLi2019, Wilde2019}, which is quite unusual in itinerant antiferromagnets.  A small amount (2.5\%) of La doping has been reported to cause FM ordering in Sr$_{1-x}$La$_x$Co$_2$As$_2$~\cite{Shen2018,Shen2019}. In \cca, the A-type AFM order was found to be smoothly suppressed by Fe doping on the Co site~\cite{Jayasekara2017}.

The above studies illustrate the impact of chemical substitution/doping on the physical properties exhibited by the Co122 systems. Since the alkaline-earth  Co-As systems exhibit itinerant magnetism that originates from the properties of band electrons near the Fermi surface, the magnetic properties strongly depend on the electronic effects of substituting/doping by different elements. Most such  studies have been carried out using 3$d$- and $4d$-transition metal substitutions on the Co-sites, and the effect of substituting $5d$ atoms for Co has not been emphasized.  Moreover, the high atomic number 5$d$ elements are expected to significantly alter the magnetic interaction in those systems due to the strong spin-orbit coupling associated with such atoms. Spin-orbit coupling has been considered to be an important tool to tune the superconducting and magnetic properties of different systems as reported earlier~\cite{Smidman2017, Marco2010, Ng2000, Rhodes2015, Plumb2014, Ma2017}. Thus, it is interesting to examine whether 5$d$-element substitutions on the Co~site can reveal novel properties and ground states.

In this work, we report the influence of Ir (5$d$) substitutions for Co in CaCo$_{1.86}$As$_2$ on the crystallographic, magnetic and thermal properties. Although all the \ccia\ compounds are found to crystallize in the ucT structure as in the  parent compound, the Ir substitutions significantly alter the magnetic interactions in these systems. A composition-induced crossover from the A-type AFM state to a magnetically-disordered FM cluster-glass state is observed. The $x = 0.14$ composition exhibits signatures of FM quantum fluctuations with a concomitant significant increase in the electronic Sommerfeld coefficient.

The experimental details are given in Sec.~\ref{Sec:ExpDet}.  The crystallography results are presented in Sec.~\ref{Sec:Cryst}, magnetization and magnetic susceptibility data in Sec.~\ref{Sec:chiM}, a study of the magnetism of the glassy state in Sec.~\ref{Sec:MagGlass}, and the heat capacity measurements in Sec.~\ref{Sec:Cp}.  A summary of the results is given in Sec.~\ref{ConcRem}.

\section{\label{Sec:ExpDet} Experimental details}

Single crystals of \ccia\ with $x$ = 0, 0.033, 0.065, 0.10, 0.14, 0.17, 0.25, and 0.35 were grown out of (Co,Ir)As self flux using the high-temperature solution-growth technique. The high-purity starting materials Ca (99.999\%, Alfa Aesar), Co (99.998\%, Alfa Aesar), Ir (99.9999\%, Ames Laboratory), and As (99.9999\%, Alfa Aesar) were taken in the molar ratio Ca:Co:Ir:As = 1:$4(1 - x$):$4x$:4 and placed in an alumina crucible. The crucible was then sealed in a silica tube under $\approx 1/4$~atm of Ar gas. Quartz wool was placed above the filled crucible to extract the flux during centrifugation. The assembly was preheated to 650~$^{\circ}$C for 6~h and then heated to 1300~$^{\circ}$C at 50~$^{\circ}$C/h. The sample was kept at that temperature for 20~h for homogenization. Then the tube was cooled to 1180~$^{\circ}$C at a rate of 6~$^{\circ}$C/h and the single crystals were separated from the flux using a centrifuge. Shiny platelike single crystals of different sizes were obtained from the growths with the $c$~axis perpendicular to the plate surfaces.  However, the crystal size and homogeneity both decreased with increasing Ir substitution, so we could not obtain crystals with $x > 0.35$.

The phase homogeneity and the average composition of the \ccia\ crystals were determined using a scanning-electron microscope equipped with an energy-dispersive x-ray spectroscopy (EDS) attachment from JEOL\@. The chemical compositions were measured at many points on both surfaces of the platelike crystals to confirm their chemical homogeneity. The average compositions of the crystals used for different measurements are listed below in Table~\ref{CrystalData}.

Single-crystal x-ray diffraction (XRD) measurements were performed at room temperature on a Bruker D8 Venture diffractometer operating at 50~kV and 1~mA equipped with a Photon 100 CMOS detector, a flat graphite monochromator and a Mo~K$\alpha$ I$\mu$S microfocus source ($\lambda = 0.71073$~\AA). The raw frame data were collected using the Bruker APEX3 program~\cite{APEX2015}, while the frames were integrated with the Bruker SAINT software package~\cite{SAINT2015} using a narrow-frame algorithm for integration of the data and were corrected for absorption effects using the multiscan method (SADABS) \cite{Krause2015}.  The atomic thermal factors were refined anisotropically.  Initial models of the crystal structures were first obtained with the program SHELXT-2014 \cite{Sheldrick2015A} and refined using the program SHELXL-2014 \cite{Sheldrick2015C} within the APEX3 software package.

The $M(H,T)$ data were obtained using a Quantum Design, Inc., magnetic-properties measurement system (MPMS) SQUID magnetometer in the range $T=1.8$ to~300~K with magnetic fields up to 5.5~T (1~T~$\equiv 10^4$~Oe). The heat capacity $C_{\rm p}(H,T)$ measurements were performed using the relaxation technique in a Quantum Design, Inc., physical-properties measurement system (PPMS) in the ranges $T=1.8$--300~K and $H = 0$--9~T\@.

\section{\label{Sec:Cryst} Crystallography}

\begin{figure}
\includegraphics[width = 2.25in]{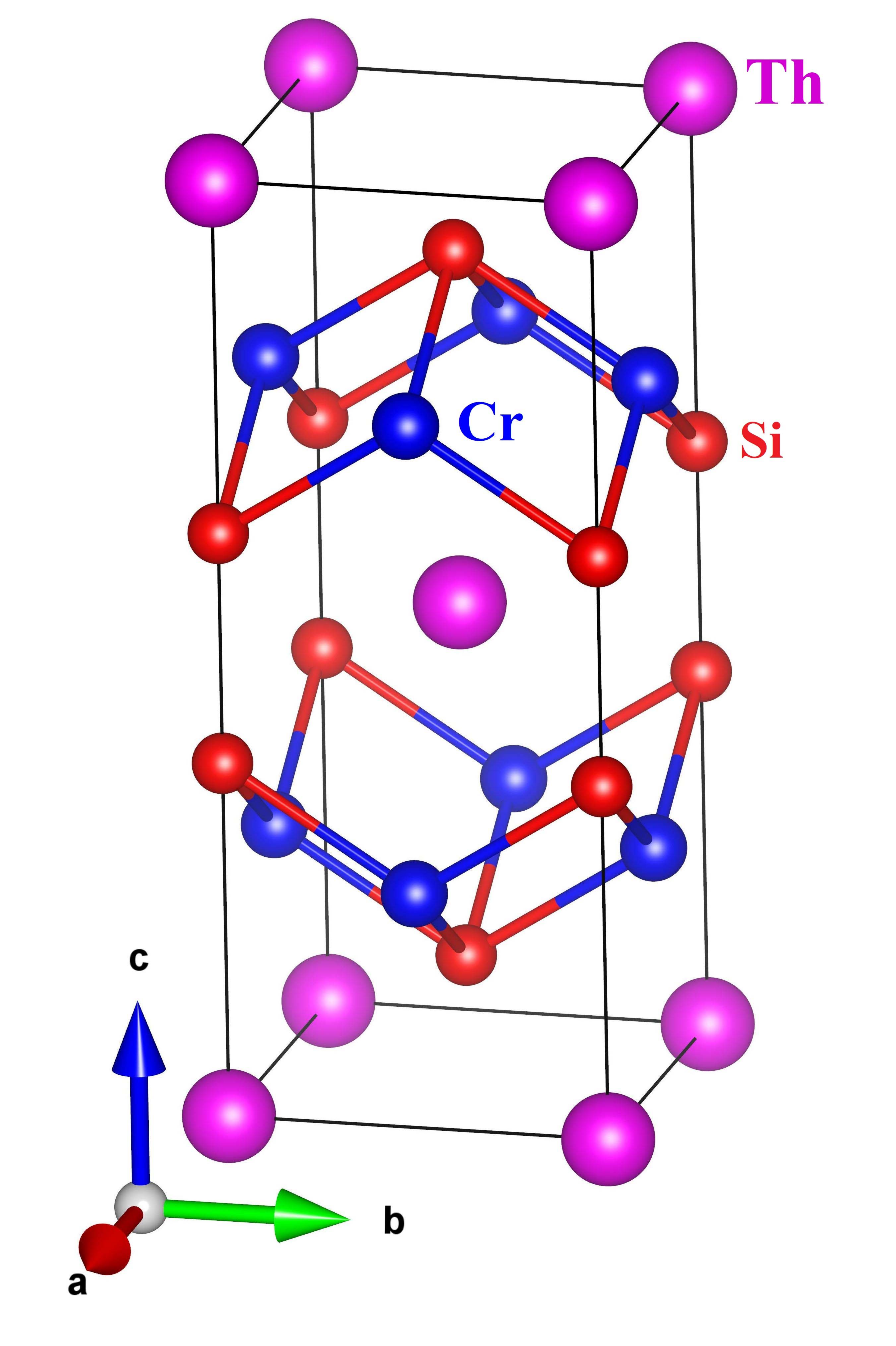}
\caption{Unit cell of the body-centered tetragonal ThCr$_2$Si$_2$ crystal structure.}
\label{Crystalstructure}
\end{figure}

\begin{figure}
\includegraphics[width = 3in]{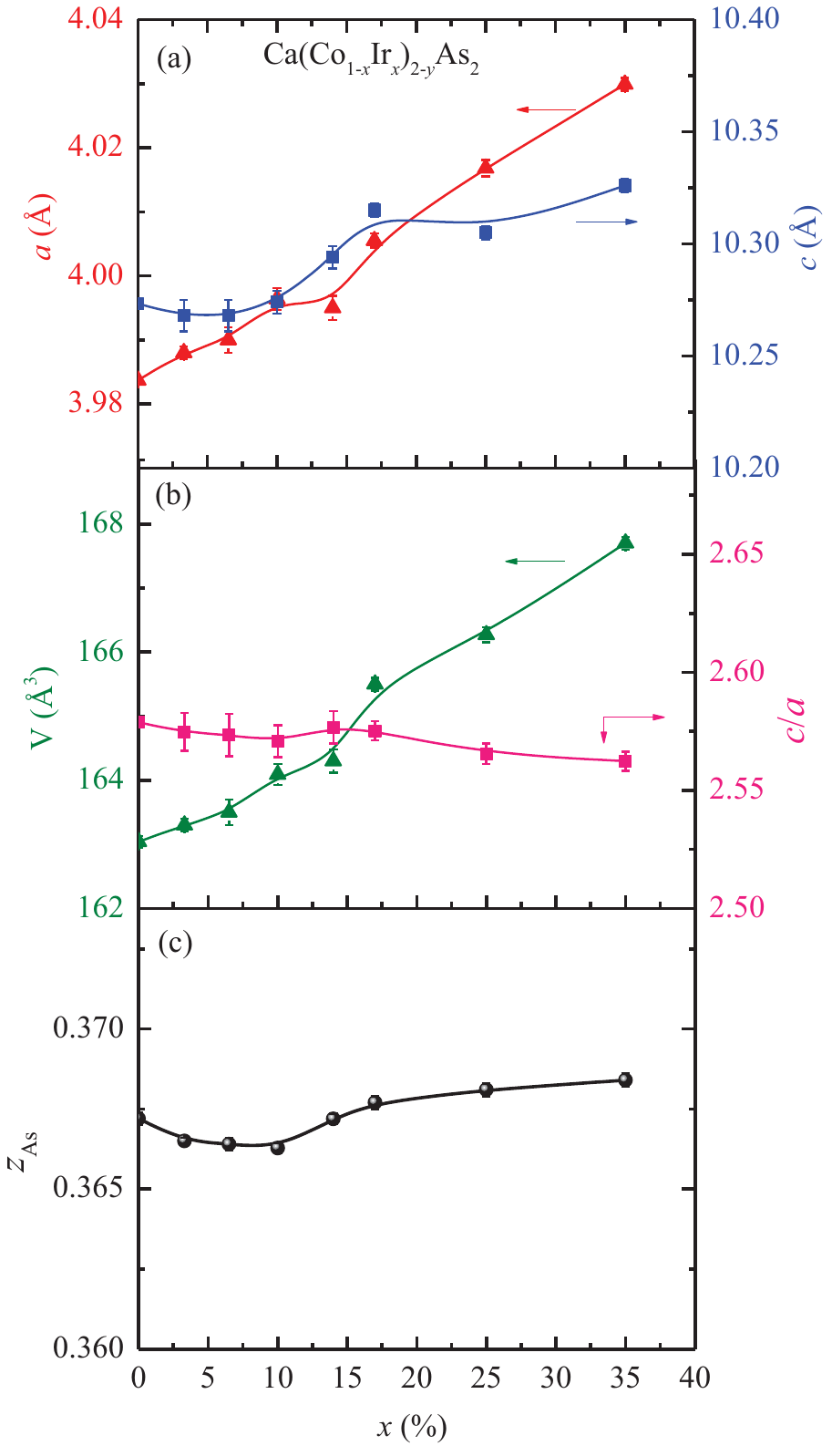}
\caption{Crystallographic parameters (a)~tetragonal lattice parameters $a$ and $c$, (b)~Unit cell volume $V$ and $c/a$ ratio, and (c)~$c$-axis As position parameter ($z_{\rm As}$) of \ccia\ compounds as a function of Ir substitution $x$.  The lines are guides to the eye.}
\label{unitcellparameters}
\end{figure}

\begin{table*}[t!]
\caption{\label{CrystalData} Room-temperature crystallographic data for \ccia\ ($0\leq x \leq 0.35,\ 0.10\leq y \leq 0.14$) single crystals. The labeled compositions were estimated from EDS analyses. Also listed are the tetragonal lattice parameters $a$ and~$c$, the unit cell volume $V_{\rm cell}$, the $c/a$ ratio, and the fractional $c$-axis position of the As site ($z_{\rm AS}$) obtained from single-crystal XRD data.}
\begin{ruledtabular}
\begin{tabular}{ cccccc }
 Compound  & $a$ (\AA)  & $c$ (\AA) & $V_{\rm cell}$ (\AA$^3$) & $c/a$ & $z_{\rm As}$\\
\hline
CaCo$_{1.86(2)}$As$_2$                               			&   3.9837(4)   	&   10.2733(4)  &   163.04(9)  &   2.5788(6)   &   0.3672(4)   \\
Ca(Co$_{0.967(3)}$Ir$_{0.033(3)}$)$_{1.86(2)}$As$_2$   &   3.988(1)    	&   10.268(7)    &   163.3(1)  &   2.575(8)   &   0.3665(1)   \\
Ca(Co$_{0.935(5)}$Ir$_{0.065(5)}$)$_{1.86(2)}$As$_2$   &   3.990(2)    	&   10.268(7)    &   163.5(2)  &   2.573(9)   &   0.3664(2)   \\
Ca(Co$_{0.90(1)}$Ir$_{0.10(1)}$)$_{1.86(2)}$As$_2$      &   3.996(2)    	&   10.274(5)    &   164.1(2)  &   2.571(7)   &   0.3662(2)   \\
Ca(Co$_{0.86(1)}$Ir$_{0.14(1)}$)$_{1.87(2)}$As$_2$      &   3.996(2)    	&   10.294(5)    &   164.3(2)  &   2.577(7)   &   0.3672(2)   \\
Ca(Co$_{0.83(1)}$Ir$_{0.17(1)}$)$_{1.87(2)}$As$_2$      &   4.005(1)    	&   10.315(3)    &   165.5(1)  &   2.575(4)   &   0.3677(2)   \\
Ca(Co$_{0.75(2)}$Ir$_{0.25(2)}$)$_{1.89(2)}$As$_2$      &   4.017(1)    	&   10.305(3)    &   166.3(1)  &   2.565(4)   &   0.3681(2)   \\
Ca(Co$_{0.65(4)}$Ir$_{0.35(4)}$)$_{1.90(2)}$As$_2$      &   4.029(1)    	&   10.326(3)    &   167.7(1)  &   2.562(4)   &   0.3684(2)   \\
\end{tabular}
\end{ruledtabular}
\end{table*}

The room-temperature single-crystal XRD measurements demonstrated that the \ccia\ ($0 \leq x \leq 0.35$) crystals form in the body-centered tetragonal ThCr$_2$Si$_2$-type crystal structure (space group $I{\rm 4}/mmm$) shown in Fig.~\ref{Crystalstructure}. The crystallographic parameters are listed in Table~\ref{CrystalData}. Generally, when the $c/a$ ratio is less than 2.67, the system is considered to form with a cT crystal structure~\cite{Anand2012}. The $c/a$ values for the present crystals are well below that value indicating that all the Ir-substituted compositions in our study form in a cT structure as does the undoped \cca\ parent compound. The EDS results reveal that the vacancy concentration on the Co site changes from 7(1)\% in the parent CaCo$_{1.86(2)}$As$_2$ compound to 5(1)\% with 35\% Ir substitution. As a single-crystal XRD refinement does not allow for simultaneous refinement of the fraction Co/Ir and total occupation of the position, only the total occupancies were refined based on the Co/Ir ratio taken from the EDS data.   The variation in the Ir content within the crystals as reflected in the compositional error bars was found to be small for low Ir substitution levels. However, the inhomogeneity increases for 35\% Ir-substituted crystals. The crystallographic parameters are plotted versus Ir concentration~$x$ in Figs.~\ref{unitcellparameters}(a)--\ref{unitcellparameters}(c).  Both lattice parameters $a$ and $c$ are found to increase nonlinearly with increasing Ir substitution giving rise also to a nonlinear increase of the unit cell volume $V_{\rm cell}$ with increasing~$x$.

\section{\label{Sec:chiM} Magnetic susceptibility}

\begin{figure}
\includegraphics[width = 3.4in]{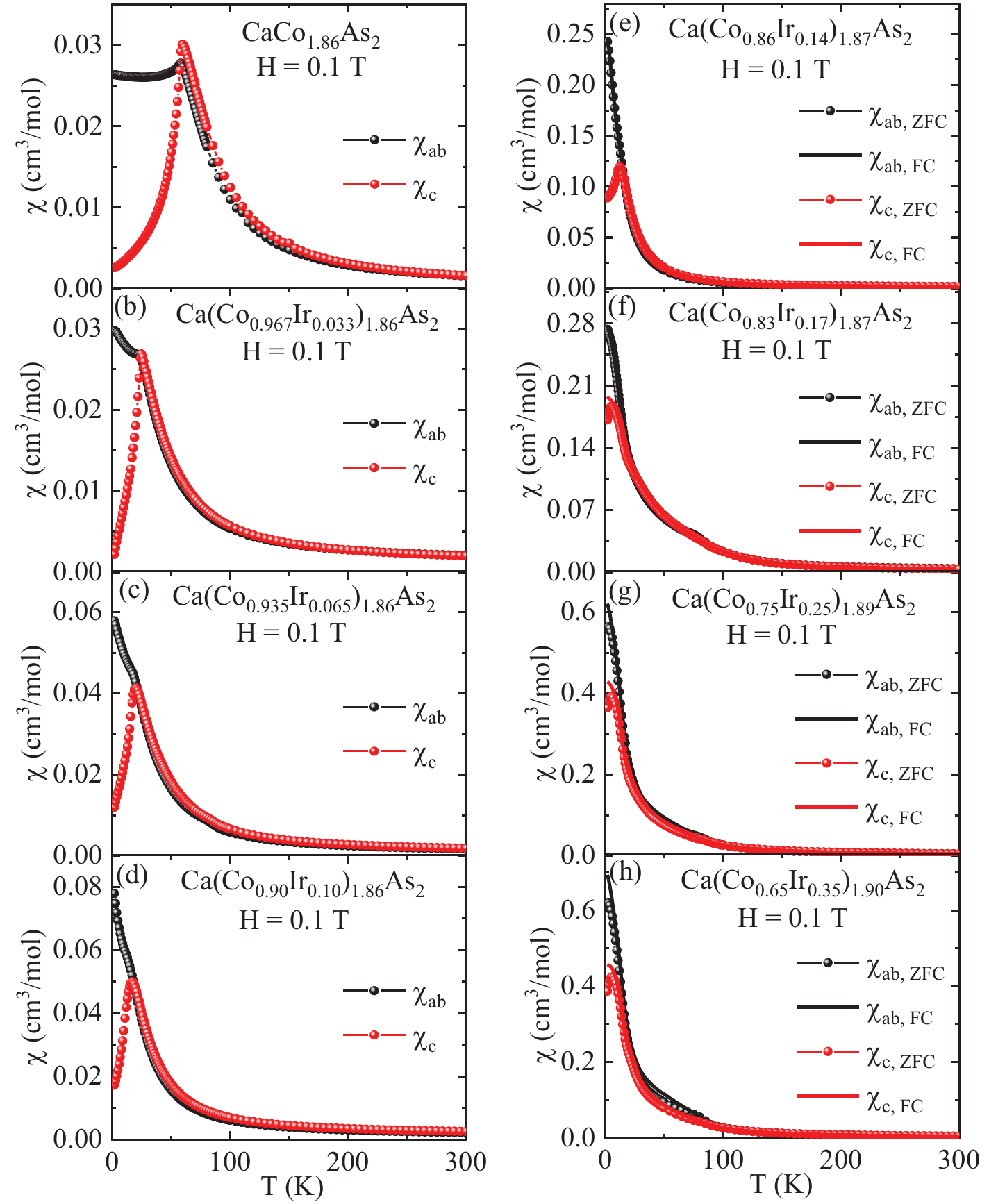}
\caption{Magnetic susceptibility ($\chi = M/H$) as a function of temperature~$T$ for \ccia\ single crystals in a magnetic field $H =0.1$~T applied in the $ab$~plane ($\chi_{ab}$) and along the $c$~axis ($\chi_c$).  Note the strong increase in the ordinate scale with increasing~$x$ and the associated increase in the temperature dependence of $\chi_c$ below the cusp temperature for $\chi_{\rm ab}$.}
\label{M-T_all_separate}
\end{figure}

The temperature dependence of the magnetization for \ccia\ crystals was measured under zero-field-cooled (ZFC) and field-cooled (FC) protocols in a magnetic field $H=0.1$~T applied in the $ab$~plane ($H~ \| ~ab$) and along the $c$~axis ($H~ \| ~c$). Figures~\ref{M-T_all_separate}(a)--\ref{M-T_all_separate}(h) show the temperature dependence of the magnetic susceptibility $\chi \equiv M/H$ for all eight \ccia\ crystals in ZFC protocol.  Evidence for some type of magnetic ordering is seen for each of the crystals. The details on the possible nature of magnetic ground states are as follows.

\subsection{Antiferromagnetic ordering temperature $T_{\rm N}$ for $0 \leq x \leq 0.14$}

\begin{figure}[h!]
\includegraphics[width = 3.2in]{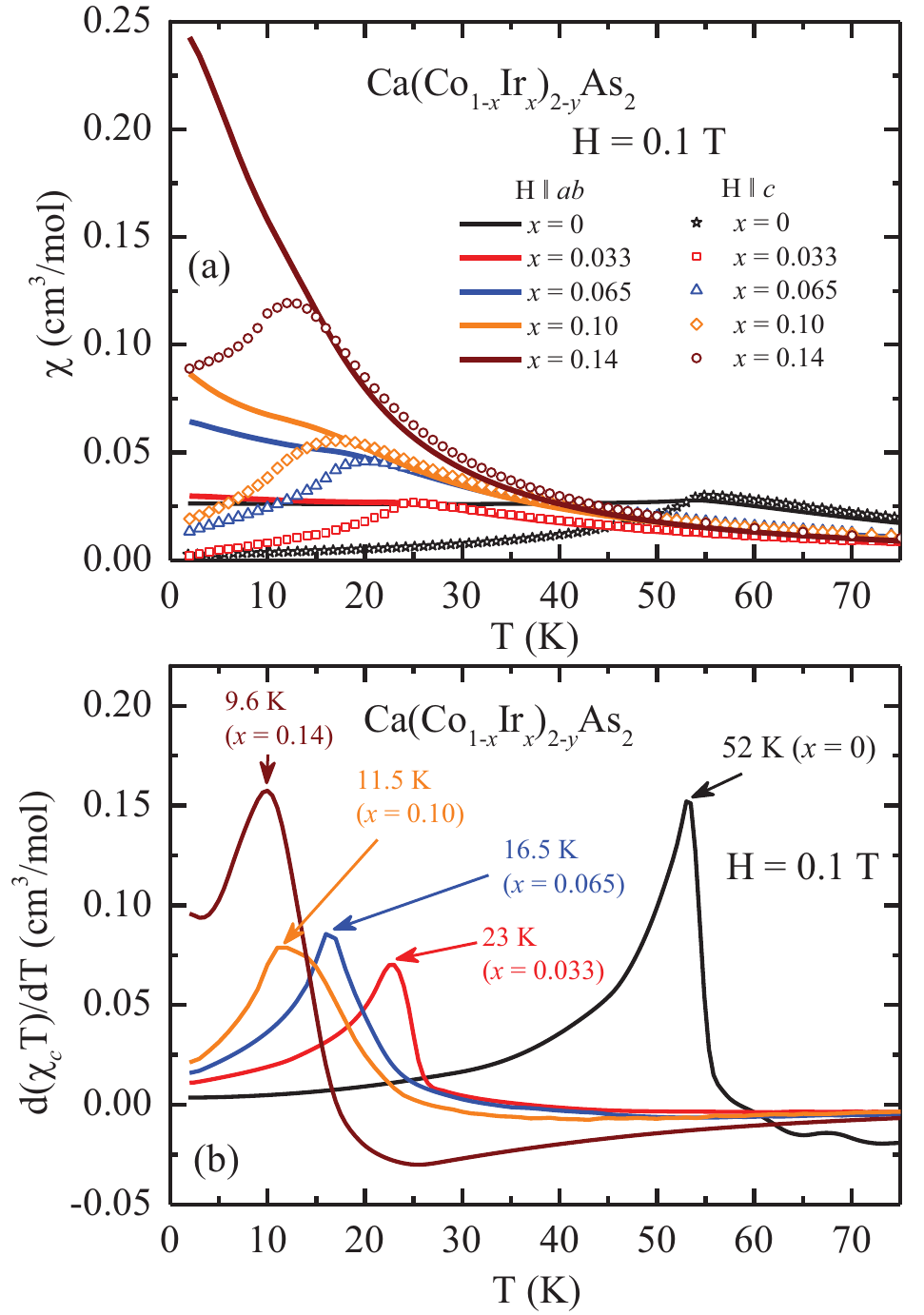}
\caption{(a)~Expanded plots of the magnetic susceptibilities $\chi_{ab}(T)$ and $\chi_c(T)$  with $H=0.1$~T for \ccia\ crystals with compositions $x=0$ to~0.14.  (b)~Temperature derivative $d(\chi_cT)/dT$ versus~$T$ obtained for the crystals, where the temperatures of the peaks are identified as the N\'eel temperatures versus Ir concentration~$x$.}
\label{M-T_all_together_0-14}
\end{figure}

The parent compound CaCo$_{1.86(2)}$As$_2$ orders in an A-type AFM structure at $T_{\rm N} = 52(1)$~K as reported earlier~\cite{Anand2014Ca}, where  $\chi_{c}(T \to 0) = 0$ and $\chi_{ab}$ is nearly independent of $T$ below $T_{\rm N}$. These are the characteristic signatures of a $c$-axis collinear antiferromagnet.  In this magnetic structure of CaCo$_{1.86(2)}$As$_2$, as noted above the ordered moments within an $ab$~plane are aligned ferromagnetically along the $c$~axis with the moments in adjacent layers along the $c$~axis aligned antiferromagnetically.  Thus, $\chi(T)$ below $T_{\rm N}$ is anisotropic with the $c$~axis as the easy axis.  Here we discuss the magnetic ground state behaviour for $x=0$--0.14.

Figure~\ref{M-T_all_together_0-14}(a) shows expanded plots of $\chi_{ab}$ and $\chi_c$ versus~$T$ for \ccia\ crystals with $x=0$ to $x=0.14$.  For antiferromagnets with an easy $c$~axis as in \cca, the N\'eel temperature is the temperature of the peak in the derivative $d(\chi_cT)/dT$~\cite{Fisher1962}.  Plots of $d(\chi_c T)/dT$ versus~$T$ are shown in Fig.~\ref{M-T_all_together_0-14}(b) where the $T_{\rm N}$ values obtained from the temperatures of the peaks are listed in Table~\ref{Tab.chidata}. The $T_{\rm N}$ decreases rapidly to $23(1)$~K for 3.3\% Ir substitution. Moreover, with increasing~$x$, the magnitudes of both $\chi_{ab}$ and $\chi_{c}$ increase significantly below $T_{\rm N}$, suggesting an increase in the FM fluctuations. Interestingly, though the $\chi_{c}$ suggests an AFM ordering below 9.6 K for $x = 0.14(1)$, $\chi_{ab}$ does not show any peak/cusp/inflection at $T_{\rm N}$ and increases monotonically for temperatures down to 2 K.

\begin{figure}
\includegraphics[width = 3.3in]{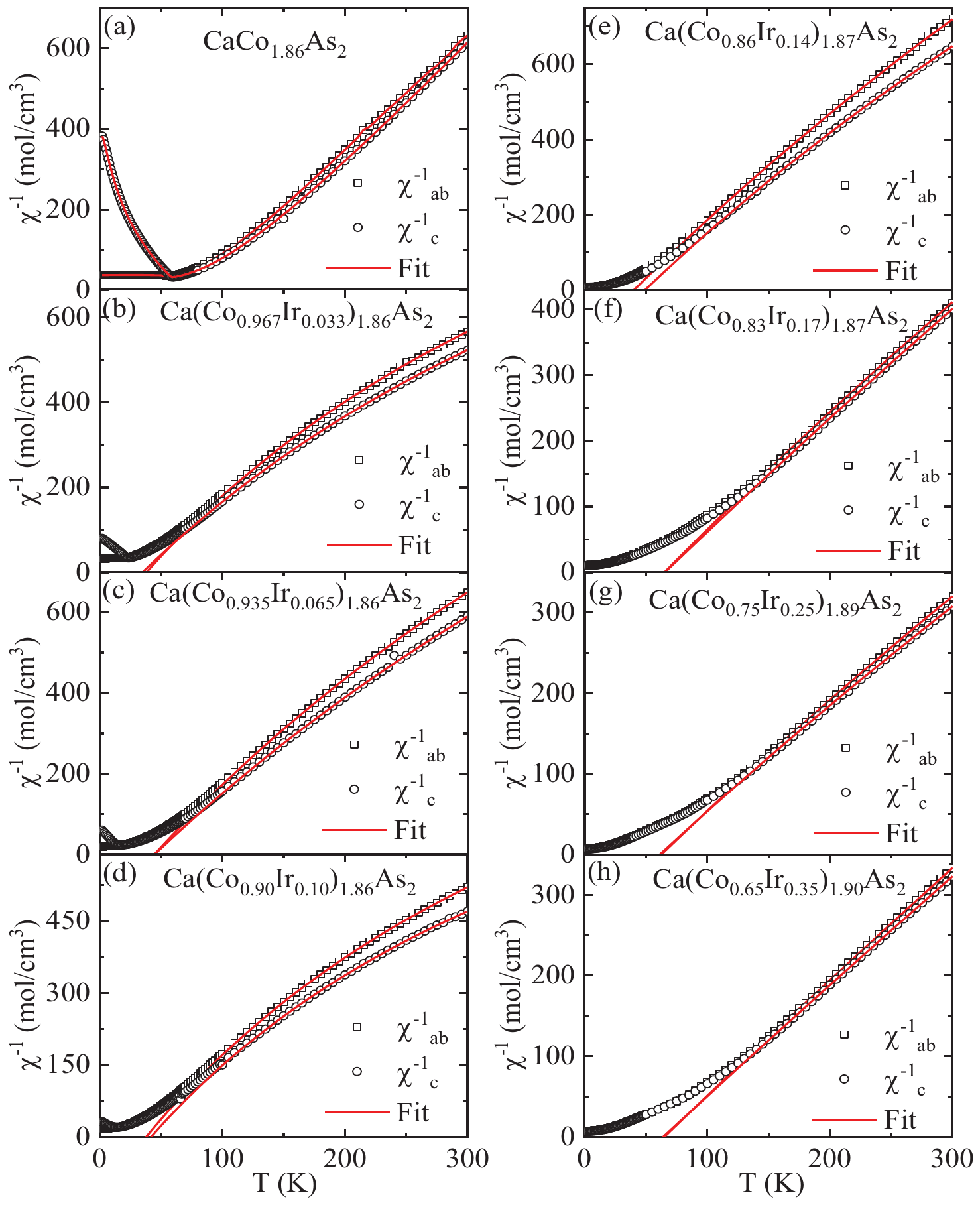}
\caption{The temperature dependence of zero-field-cooled (ZFC) inverse magnetic susceptibility in magnetic field $H = 1$~T applied along both $ab$ plane and the $c$~axis. The red solid lines are fits by the modified Curie-Weiss law in Eq.~(\ref{Eq.ModCurieWeiss}).}
\label{Chi-1_T_All}
\end{figure}

\subsection{Magnetic susceptibility in the paramagnetic state}

\begin{table*}[ht!]
\caption{\label{Tab.chidata} Parameters obtained from Modified Curie-Weiss fits to $\chi^{-1}(T)$ data between 150 and 300~K for \ccia\ by using Eq.~(\ref{Eq.ModCurieWeiss}). Shown are  the $T$--independent contribution to the susceptibility $\chi_0$,  Curie constant per mol $C_\alpha$ in $\alpha = ab, c$ directions, and Weiss temperature $\theta\rm_{p\alpha}$. The effective moment per transition metal $\mu{\rm_{eff\alpha}(\mu_B/f.u.)} = \sqrt{8C_\alpha}$ calculated from Eq.~(\ref{Eq.mueff}).  Also included are the N\'eel temperatures $T_{\rm N}$ obtained from the temperatures of the cusps in $d(\chi_{c}T)/dT$ in Fig.~\ref{M-T_all_together_0-14}(b), the $ab$-plane spin-flop fields $H_{\rm SF}$ at $T=2$~K, and the blocking temperatures $T_{\rm B}$. }
\begin{ruledtabular}
\begin{tabular}{ccccccccc}	
  										& Field			& $\chi_0$ 	& $C_{\alpha}$ &  $\mu_{\rm eff\alpha}$ 	& $\theta_{\rm p\alpha}$ & $T_{\rm N}$ & $H_{\rm SF}$ & $T_{\rm B}$  \\
Compound 								& Orientation		& $\rm{\left(10^{-4}~\frac{cm^3}{mol}\right)}$	 & $\rm{\left(\frac{cm^3 K}{mol}\right)}$    & $\rm{\left(\frac{\mu_B}{mol}\right)}$& (K)  & (K) & (kOe) & (K)\\
\hline

CaCo$_{1.86}$As$_2$                                        	& $H\parallel ab$ 	& 0.03(2) 		& 0.354(7) 	& 1.68(2)		& 	76(1)		& 	53	&	35.0(5)	\\
						                            	& $H\parallel c$ 	& $-0.2(2)$ 	& 0.416(3) 	& 1.82(1) 		&	75(1)		& 		&			\\
Ca(Co$_{0.967}$Ir$_{0.033}$)$_{1.86}$As$_2$    	& $H\parallel ab$ 	& 5.8(2)		& 0.308(8) 	& 1.57(2) 		&	38(2)		& 	23	&	22.5(5)	\\	
							                   	& $H\parallel c$ 	& 5.72(6)		& 0.355(2) 	& 1.67(1) 		&	34.8(5)	& 		&			\\
Ca(Co$_{0.935}$Ir$_{0.065}$)$_{1.86}$As$_2$        & $H\parallel ab$ 	& 3.58(3)		& 0.316(4)		& 1.59(1)		&	44.2(2)	& 	16.5	&	17.5(1)	\\
							                  	& $H\parallel c$ 	& 3.39(5)		& 0.357(4)		& 1.69(1)		&	44.5(5)	& 		&			\\
Ca(Co$_{0.90}$Ir$_{0.10}$)$_{1.86}$As$_2$       	& $H\parallel ab$ 	& 6.9(1)		& 0.320(3)		& 1.60(1)		&	38(1)		& 	11.5	&	15.0(2)	\\
										& $H\parallel c$ 	& 7.9(1)		& 0.361(5)		& 1.70(1)		&	42(1)		& 		&			\\
Ca(Co$_{0.86}$Ir$_{0.14}$)$_{1.87}$As$_2$ 		& $H\parallel ab$ 	& 1.9(2)		& 0.310(7)		& 1.57(2)		&	40(2)		& 	9.6	&	5.0(1)	\\		
										& $H\parallel c$ 	& 2.61(5)		& 0.322(2)		& 1.60(1)		&	49(1)		& 		&			\\
Ca(Co$_{0.83}$Ir$_{0.17}$)$_{1.87}$As$_2$ 		& $H\parallel ab$ 	& 1.8(2)		& 0.529(8)		& 2.06(2)		&	65(1)		& 		&			&	6.1	\\
										& $H\parallel c$ 	&1.1(2) 		& 0.559(7)		& 2.11(1)		&	65.4(9)	& 		&			\\
Ca(Co$_{0.75}$Ir$_{0.25}$)$_{1.89}$As$_2$ 		& $H\parallel ab$	& 2.4(2)		& 0.686(6)		& 2.34(1)		&	62.3(7)	& 		&			&	5.5	\\
										& $H\parallel c$ 	& 2.5(4)		& 0.716(12)	& 2.39(2)		&	61(1)		& 		\\
Ca(Co$_{0.65}$Ir$_{0.35}$)$_{1.90}$As$_2$ 		& $H\parallel ab$ 	& 1.0(2)		& 0.681(7)		& 2.35(1)		&	65(1)		& 		&			&	2.5	\\
										& $H\parallel c$ 	& 0.8(2)		& 0.715(11)	& 2.39(2)		& 	64(1)		& 		\\
\end{tabular}
\end{ruledtabular}
\end{table*}

\begin{figure}[h!]
\includegraphics[width = 2.3in]{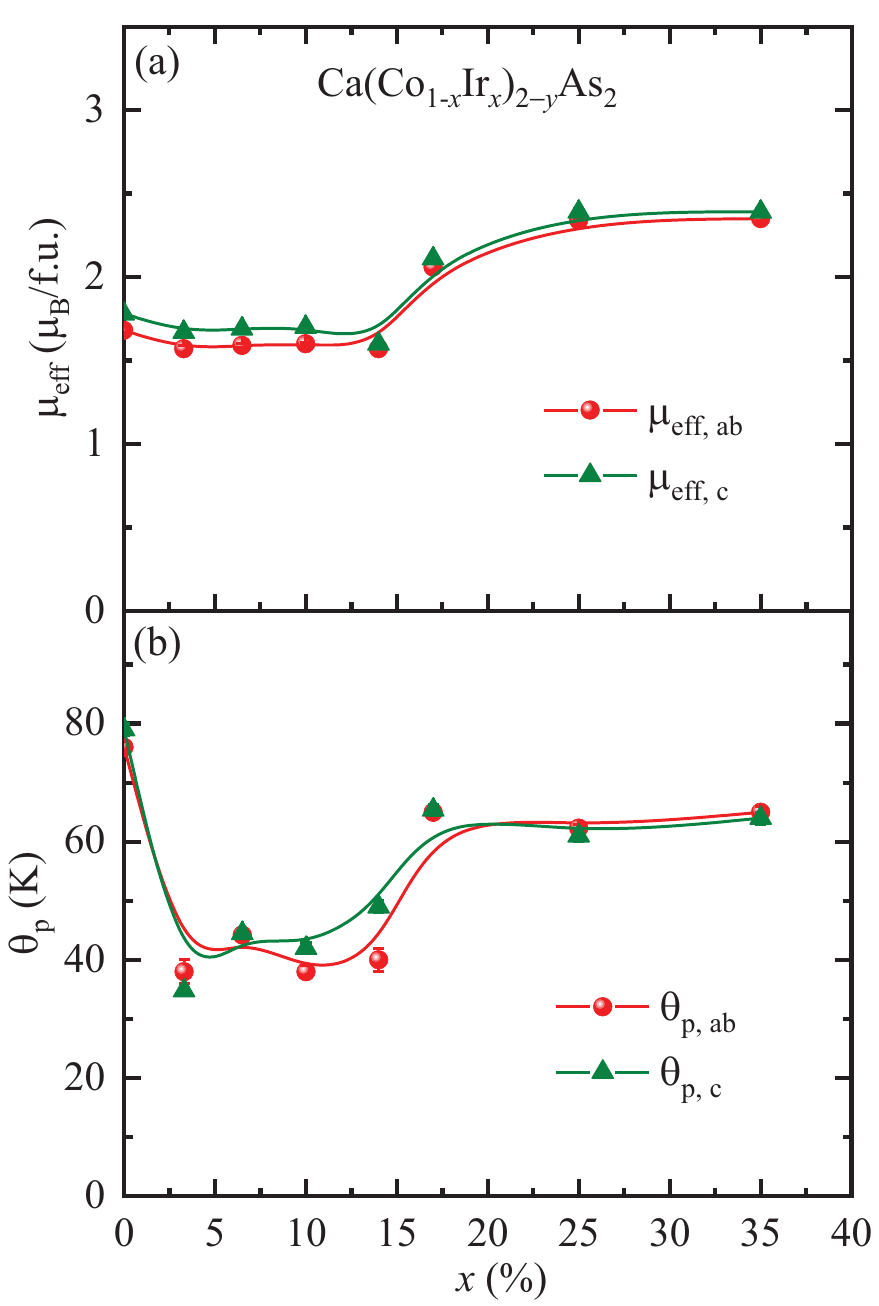}
\caption{Effective magnetic moment $\mu_{\rm_{eff}}$ and the Weiss temperature $\theta_{\rm p}$ as a function of $x$. The lines are guides to the eye.}
\label{M-T_parameters}
\end{figure}

The $\chi(T)$ data in the paramagnetic (PM) state at $T>T_{\rm N}$ are analyzed in terms of local moments using the modified Curie-Weiss law
\bea
\chi_{\alpha}(T) =\chi_0 + \frac{C_{\alpha}}{T-\theta_{\rm p\alpha}} \qquad (\alpha = ab,\ c),
\label{Eq.ModCurieWeiss}
\eea
where $\chi_0$ is an isotropic temperature-independent term that contains the diamagnetic contributions from the atomic cores and the conduction-carrier orbital Landau susceptibilities, together with the paramagnetic contribution from the Pauli spin susceptibility  of the conduction carriers. The Curie constant per mole of formula units~(f.u.) is given by
\bea
C_{\alpha}&=&\frac{N_{\rm A} {g_\alpha}^2S(S+1)\mu^2_{\rm B}}{3k_{\rm B}} = \frac{N_{\rm A}\mu^2_{\rm {eff \alpha}}} {3k_{\rm B}} \quad (\alpha=ab,\ c),  \nonumber\\
\label{Eq.Cvalue1}
\eea
where $N_{\rm A}$ is Avogadro's number, $g_\alpha$ is the spectroscopic splitting factor ($g$ factor), $S$ is the spin angular-momentum quantum number, $k_{\rm B}$ is Boltzmann's constant,  and $\mu_{\rm eff}$ is the effective moment of a spin in units of Bohr magnetons $\mu\rm_B$.
Inserting the Gaussian cgs values of the fundamental constants into Eq.~(\ref{Eq.Cvalue1}), the Curie constant per mole of spins is expressed as
\bea
C_{\alpha} {\rm (cm^3 \,K/mol)} \approx \frac{\mu^2_{\rm eff}}{8} ~(\mu{\rm_ B/f.u.}).
\label{Eq.Cvalue2}
\eea
Hence
\bea
\mu_{\rm eff \alpha}~ {\rm (\mu_B/f.u.)} \approx \sqrt{8C_{\alpha}}.
\label{Eq.mueff}
\eea

The anisotropic inverse susceptibilities of the \ccia\ crystals are plotted versus temperature in Fig.~\ref{Chi-1_T_All}. The fitted parameters for the $\chi^{-1}(T)$ data in Fig.~\ref{Chi-1_T_All} obtained over the temperature range 150~K to 300~K using the modified Curie-Weiss law are listed in Table~\ref{Tab.chidata}.   The fits are shown as the solid red curves in Fig.~\ref{Chi-1_T_All}.  The effective moment $\mu_{\rm_{eff}}$ and Weiss temperature $\theta_{\rm_{p}}$ are plotted versus~$x$ in Figs.~\ref{M-T_parameters}(a) and~\ref{M-T_parameters}(b), respectively.

For spins~$S$ with $g=2$, the isotropic Curie constant in units of ${\rm cm^3\,K/mol\,spins}$ is
\bea
C = 0.5002 S(S+1).
\label{Eq.Curieconstant}
\eea
The \ccia\ system has approximately 1.9~transition-metal atoms per formula unit and assuming that the Co/Ir atoms carry a local moment,  Eq.~(\ref{Eq.Curieconstant}) gives
\bea
C_{\rm mol} \approx 0.95 S(S+1)
\label{Eq.CCoIr}
\eea
per mole of \ccia\ formula units.  Thus for $S=1/2$ with $g=2$ we expect $C_{\rm mol} \approx 0.74~{\rm cm^3\,K/mol}$, whereas $S=1$  gives  $C_{\rm mol} \approx 1.80~{\rm cm^3\,K/mol}$.  From Table~\ref{Tab.chidata}, the Curie constants are in the range 0.31--0.36~$\rm{cm^3\,K/mol}$ for $0 \leq x \leq 0.14$ and 0.53--0.72~$\rm{cm^3\,K/mol}$ for $0.17\leq x \leq 0.35$.  Thus there is a significant change in the magnetic character of  \ccia\ between the composition ranges $0\leq x \leq 0.14$ and $0.17 \leq x \leq 0.35$, suggesting the presence of a magnetic phase boundary at $x\approx 0.15$.  In addition, the large discrepancy between the measured values for $x\leq 0.14$ and the values obtained from Eq.~(\ref{Eq.CCoIr}) suggests  that the magnetism is itinerant for these compositions as previously deduced for $x=0$ (see, e.g., \cite{Sapkota2017}).  On the other hand, the larger values of $C_{\rm mol}$ for the range  $0.17 \leq x \leq 0.35$ compared with  those at lower~$x$ values suggests an increased local-moment character for $x\geq 0.17$.

\begin{figure}
\includegraphics[width = 3.5in]{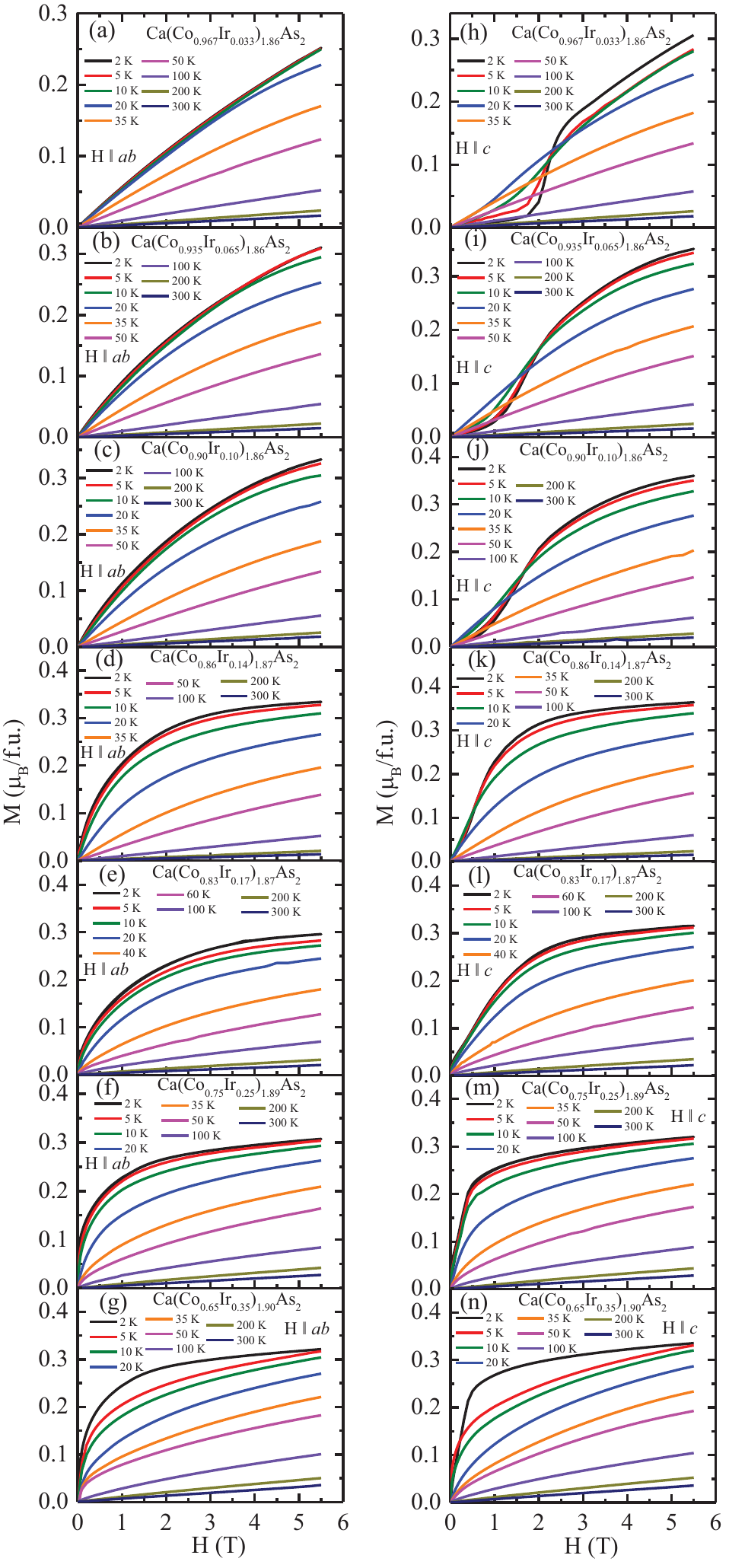}
\caption{(a)--(g)~Magnetic field dependence of in-plane isothermal magnetization ($M_{ab}$) measured at different temperatures for the \ccia\ crystals when $H \parallel ab$. (h)--(n)~Out-of-plane isothermal magnetization ($M_{c}$) versus~$H$ measured at different temperatures for $H \parallel c$. Here the data have been shown for the crystals with $x > 0$, whereas the data for $x = 0$ are given in Ref.~\cite{Anand2014Ca}.}
\label{M-H_all_separate}
\end{figure}

\subsection{Magnetization versus applied magnetic field isotherms}

\begin{figure}
\includegraphics[width = 3.3in]{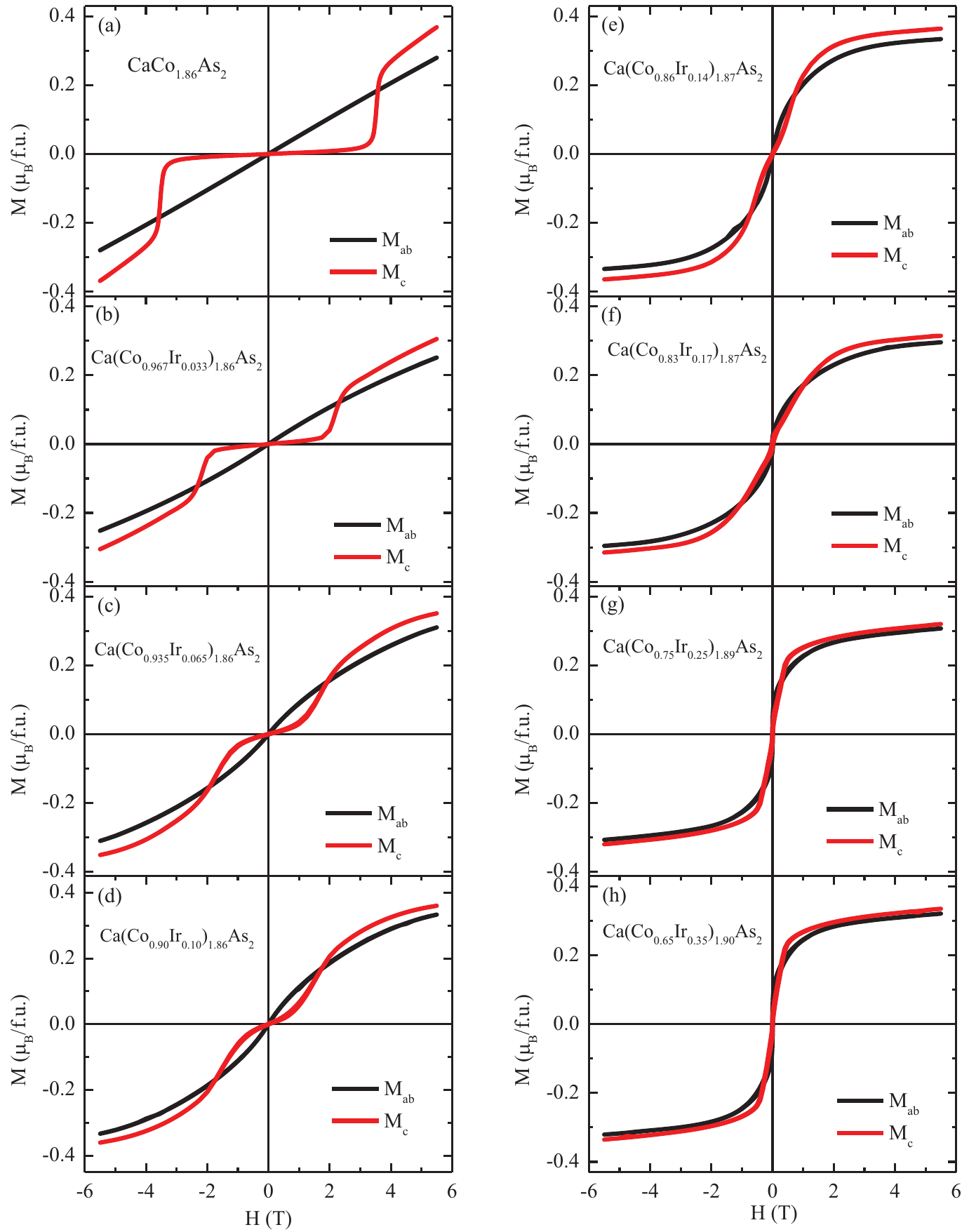}
\caption{Magnetization $M$ versus applied magnetic field~$H$ measured at $T = 2$~K for the \ccia\ crystals in the $ab$ plane ($M_{ab}$) and along $c$ axis ($M_{c}$) over the full field range of the measurements.  The $M_c(H)$ data for $x=0-0.14$ in (a)--(e) exhibit spin-flop transitions from the $c$ axis to the $ab$~plane, whereas the data for $x=0.17$--0.35 in (f)--(h) suggest the onset of an ordered FM component in the $ab$~plane. }
\label{MH_2K_all_separate}
\end{figure}

\begin{figure*}
\includegraphics[width = 6.5in]{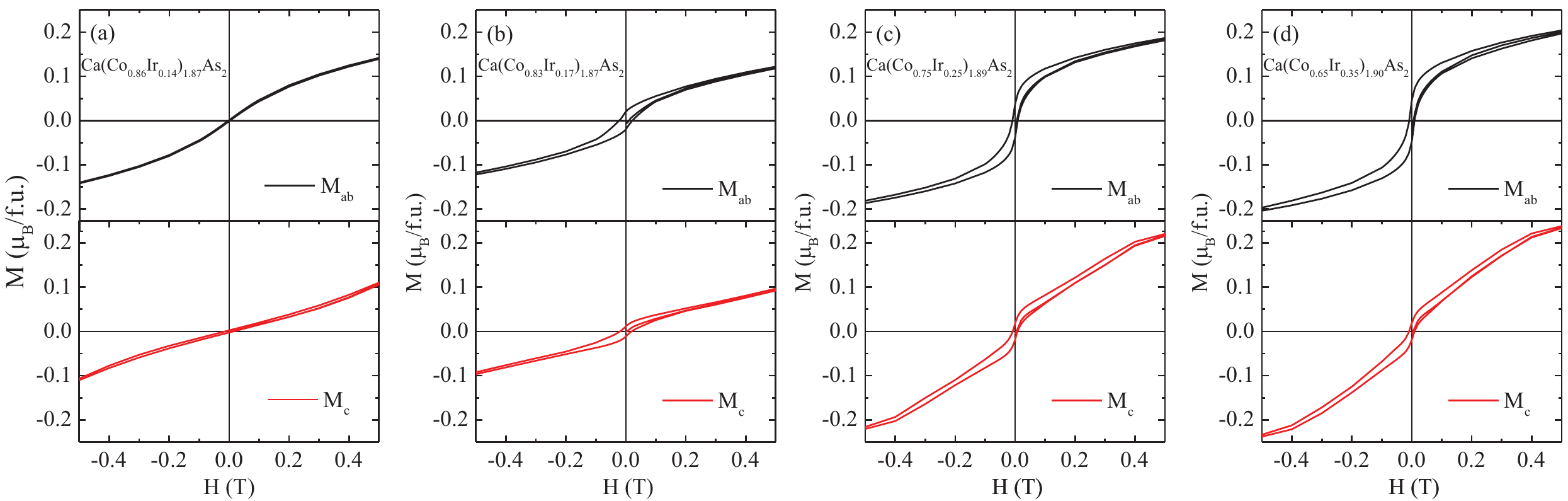}
\caption{Low field region of magnetic hysteresis behaviour measured at $T = 2$~K in the $ab$ plane (upper panel) and along $c$ axis (lower panel) for the \ccia\ crystals with (a) $x = 0.14$, (b) $x = 0.17$, (c) $x = 0.25$, and (d) $x = 0.35$.}
\label{2K_mag_hysteresis}
\end{figure*}

\begin{figure}
\includegraphics[width = 3in]{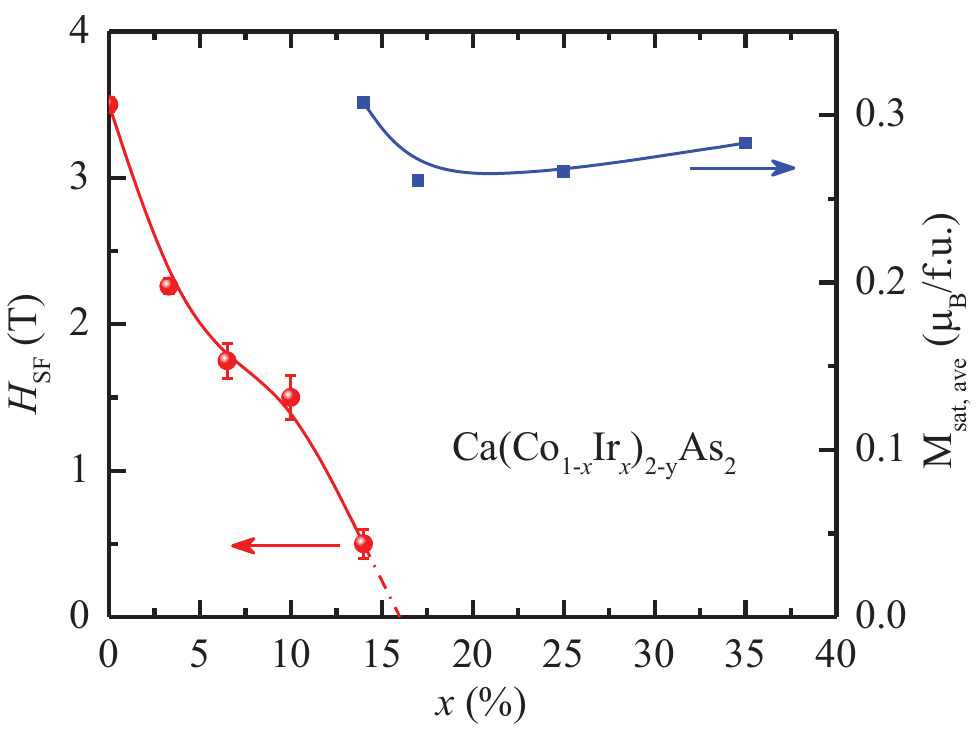}
\caption{Composition dependence of spin-flop field $H_{\rm SF}$ and spherically averaged saturation moment (M$_{\rm {sat, ave}}$) from Fig.~\ref{MH_2K_all_separate} for \ccia\ crystals with an A-type AFM ground state. The lines are guides to the eye. According to the extrapolated $H_{\rm SF}(x)$ data, the A-type AFM phase boundary is at $x\approx 0.16$.}
\label{M-H_parameters}
\end{figure}

 To further clarify the magnetic ground states in \ccia , $M(H)$ isotherms were measured at different temperatures in the field range 0--5.5 T, as shown in Figs.~\ref{M-H_all_separate}(a)--\ref{M-H_all_separate}(g) for $H \parallel ab$ and Figs.~\ref{M-H_all_separate}(h)--\ref{M-H_all_separate}(n) for $H \parallel c$. The $M(H)$ behavior in both field directions is nonlinear up to a much higher temperature than their characteristic temperature $T_{\rm N}$ or $T_{\rm B}$ due to the presence of short-range magnetic interactions. Figures~\ref{MH_2K_all_separate}(a)--\ref{MH_2K_all_separate}(h) show $M(H)$ isotherms at $T=2$~K for different compositions with both field directions. Here $M_{ab}(H)$ for $x = 0$ increases linearly with $H$ whereas $M_{c}(H)$ clearly exhibits a spin-flop (SF) transition at  $H_{\rm SF} = 3.5$~T, similar to the results reported earlier~\cite{Anand2014Ca}.   However, the SF transition field  rapidly decreases to $H_{\rm SF} = 2.15(5)$~T for $x = 0.033$, while retaining a linear $M_{ab}(H)$ behavior up to $H = 5.5$~T\@.  For both crystals, no tendency towards saturation was observed in either $M_{ab}(H)$ or $M_{c}(H)$  isotherms in the field region studied. The $H_{\rm SF}$ decreases further with increasing~$x$ and becomes negligible for $x = 0.14$. The jump in $M_{c}(H)$ at $H_{\rm SF}$  is also broadened with increasing~$x$\@.

The $M_{ab}(H)$ and $M_{c}(H)$ for $x = 0.065$ and~0.10 exhibit a saturation tendency at higher fields suggesting a reduction of AFM interactions with increasing Ir concentration. The magnetic saturation tendency is observed at much lower $H$ for $x = 0.14$ and at the highest applied field the magnetization almost saturates to values $M_{ab}^{\rm sat} = 0.33~\mu_{\rm B}$/f.u.\ and $M_c^{\rm sat}= 0.36~\mu_{\rm B}$/f.u.\  Thus the ordered moment per Co/Ir atom is only about 0.16 and 0.18~$\mu_{\rm B}$ for these two compositions.  The compounds with $x \geq 0.17$ do not show a SF transition.  Instead, $M_{ab}(H)$ and $M_{c}(H)$ almost saturate for $H \geq 3.75(10)$~T\@.

The $c$-axis spin-flop transition field  $H_{\rm SF}$ for $x=0$--0.14 and the approximate spherically-averaged saturation moment M$_{\rm {sat, ave}}$ for the \ccia\ crystals with $x=0.14$ to 0.35 are summarized in Fig.~\ref{M-H_parameters}.

\begin{table}
\caption{\label{Tab.remcoercive} Remanent magnetization ($M_{\rm rem}$) and coercive field ($H_{\rm cf}$) of \ccia\ compounds with $x = 0.17$, 0.25, and~0.35.}
\begin{ruledtabular}
\begin{tabular}{|cccc|}	
Crystal								& $H$ direction		& $M_{\rm rem}$	 	&  $H_{\rm cf}$  \\
Composition							&				&  ($\mu_{\rm B}$/f.u.)	& (Oe) \\
\hline
Ca(Co$_{0.83}$Ir$_{0.17}$)$_{1.87}$As$_2$	& $H \parallel ab$	& 0.019(2)	 &  220(2) \\
                                           				& $H \parallel c$	& 0.012(2) &  190(2) \\
\hline
Ca(Co$_{0.75}$Ir$_{0.25}$)$_{1.89}$As$_2$	& $H \parallel ab$	& 0.038(1)	 &  87(1) \\
                                           				& $H \parallel c$	& 0.018(1)	 &  88(1) \\
 \hline
Ca(Co$_{0.65}$Ir$_{0.35}$)$_{1.90}$As$_2$	& $H \parallel ab$	& 0.047(1)	 &  90(2) \\
                                           				& $H \parallel c$	& 0.019(1)	 &  92(2) \\
\end{tabular}
\end{ruledtabular}
\end{table}

No magnetic hysteresis is observed for $x \leq 0.14$.  However, the more Ir-rich compounds exhibit low-temperature magnetic hysteresis with finite coercive fields $H_{\rm cf}$ signifying a considerable increase in the FM volume fraction in these crystals as depicted in Figs.~\ref{2K_mag_hysteresis}(a)--\ref{2K_mag_hysteresis}(d) for both field directions.  The  remanent magnetization ($M_{\rm rem}$) and $H_{\rm cf}$ of the \ccia\ crystals with $x =$ 0.17, 0.25, and 0.35 are listed in Table~\ref{Tab.remcoercive}. The $H_{\rm cf}$ is maximum for $x = 0.17$, whereas it is about a factor of two smaller for $x = 0.25$ and 0.35. On the other hand, the $M_{\rm rem}$ values initially increase with increasing~$x$, where $M_{\rm rem}$ in the $ab$ plane is about a factor of two larger than along the $c$~axis. This observation correlates with the magnetic susceptibility data in Fig.~\ref{M-T_all_separate} where the $ab$-plane FM fluctuations are generally stronger than along the $c$~axis.

\section{\label{Sec:MagGlass} Magnetism of the glassy state for $0.17 \leq x \leq 0.35$}

\begin{figure*}
\includegraphics[width = 6.9in]{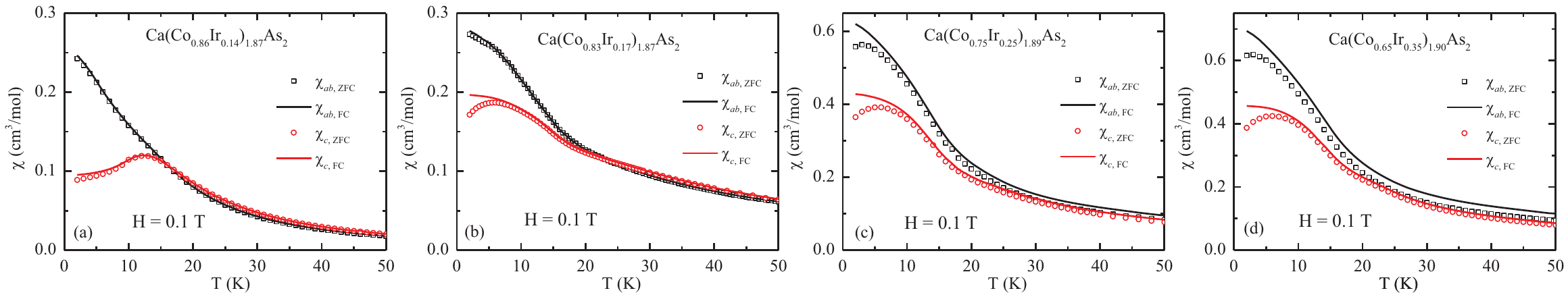}
\caption{Expanded plots at low temperatures of the magnetic susceptibilities $\chi_{ab}(T)$ and $\chi_c(T)$  under ZFC and FC conditions with $H=0.1$~T for \ccia\ crystals with compositions $x=0.14$ to~0.35.}
\label{M-T_all_together_14-35}
\end{figure*}

\begin{figure}
\includegraphics[width = 3.in]{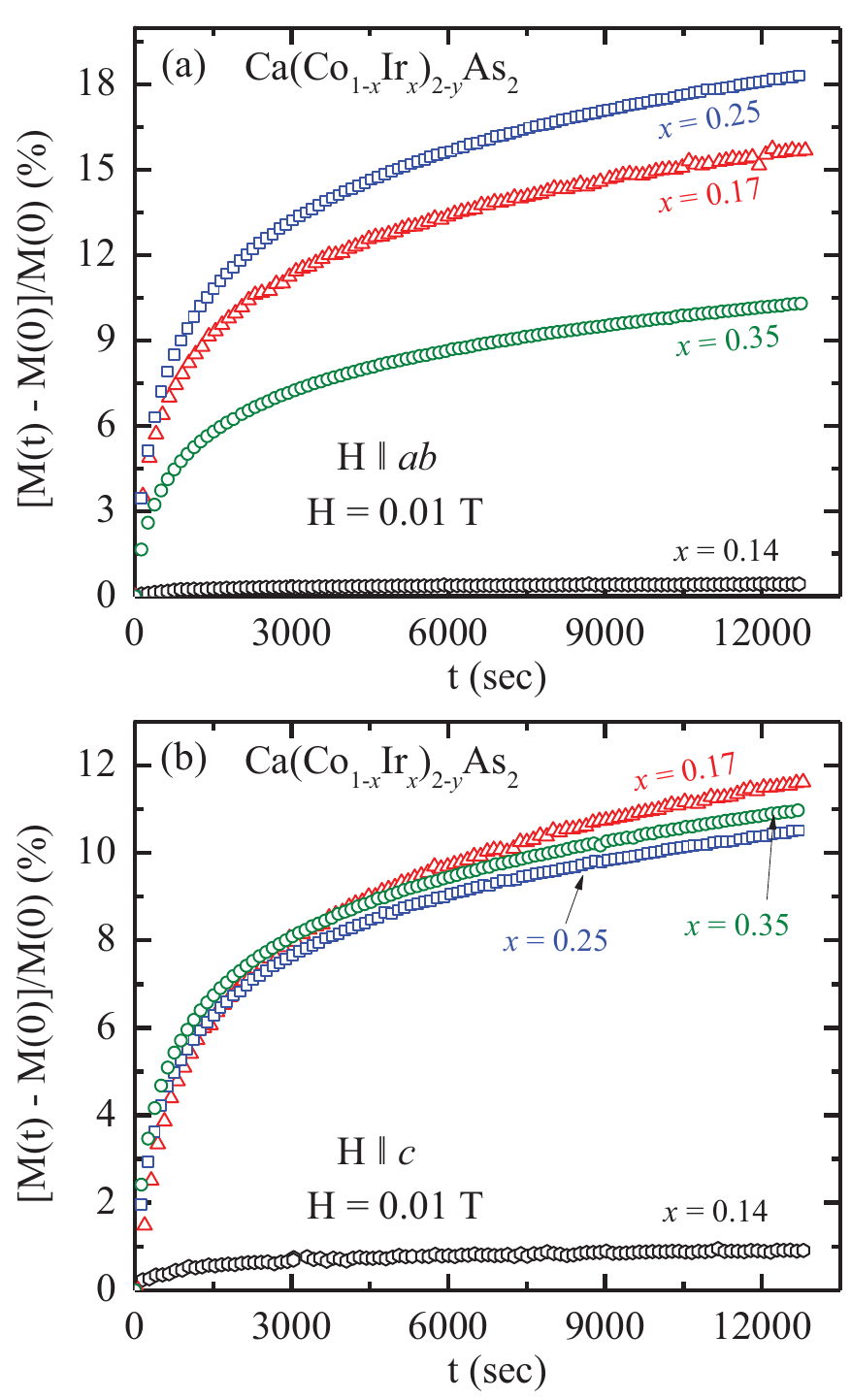}
\caption{Relative magnetic relaxation $(M(t) - M(0)/M(0)$  versus time~$t$ of \ccia\ crystals with $x \geq 0.14$ at $T = 2$~K when a small field $H = $ 0.01 T is applied at $t=0$ (a)~parallel to the $ab$ plane and (b)~along the  $c$~axis.}
\label{Relaxation_compare}
\end{figure}

\begin{figure}
\includegraphics[width = 3.4in]{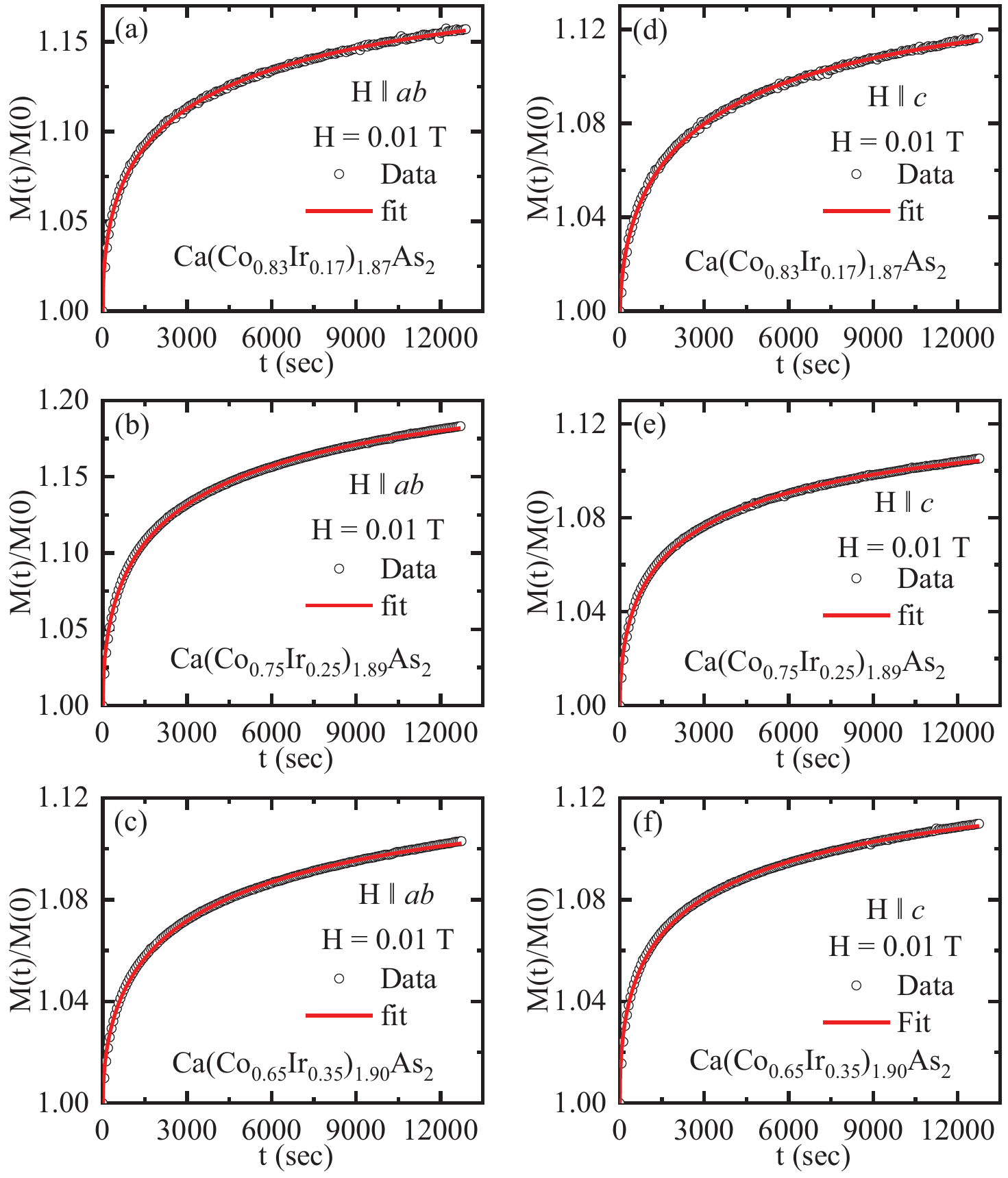}
\caption{(a)--(c) In-plane ($H \parallel ab$) and (d)--(f) out-of-plane ($H \parallel c$) magnetic relaxation $M(t)/M(0)$ versus time~$t$ for $x = 0.17$, 0.25, and 0.35 crystals, respectively, along with stretched exponential fits by  Eq.~(\ref{Eq:Mrelax}). }
\label{Relaxation_fit}
\end{figure}

Figure~\ref{M-T_all_together_14-35} depicts $\chi_{ab}(T)$ and $\chi_{c}(T)$ for $x=0.14$, 0.17, 0.25, and 0.35 measured in $H = 0.1$~T, each under both ZFC and FC conditions. The data for the AFM composition $x = 0.14$ show no significant hysteresis between the FC and ZFC curves of the respective $\chi_{ab}(T)$ and $\chi_{c}(T)$ measurements.  On the other hand, for the compositions $x = 0.17$, 0.25, and 0.35, the $\chi_{c}$ data for each~$x$ exhibit a broad maximum at a temperature denoted the blocking temperature $T_{\rm B}$ as listed in Table~\ref{Tab.chidata}, and $\chi_{ab}$ shows a FM-like saturation tendency below $T_{\rm B}$. The hysteretic behavior of the $x = 0.17$, 0.25, and 0.35 compounds of the respective $\chi_{ab}(T)$ and $\chi_{c}(T)$ measurements is similar to the corresponding behavior for different glassy systems consisting FM clusters~\cite{Pakhira2018, DXLi2003, Freitas2001, Nam1999, Anand2012PrRhSn3}. In a spin-glass system spin freezing occurs below a blocking temperature $T_{\rm B}$ (often called $T_{\rm f}$). Although the measurement of $T_{\rm B}$ is best estimated through ac magnetic susceptibility measurements, the $T_{\rm B}$ found from that measurement is the same temperature as the temperature of the maximum in the ZFC dc magnetic susceptibility measurement~\cite{Mydosh1993,Binder1986}.  Thus, the data in Fig.~\ref{M-T_all_together_14-35} for $x = 0.17$, 0.25, and 0.35 are consistent with formation of FM clusters in these crystals.

A magnetic glassy state is metastable and found to exhibit a time-dependent relaxation behavior~\cite{Mydosh1993, Nordblad1986, Chu1994}. Magnetic relaxation dynamics can be studied in different ways. In the present work, the crystals were cooled to 2~K from 300~K in zero applied field by quenching the superconducting magnet in the magnetometer before cooling.  After temperature stabilization at 2~K, a small magnetic field $H = 0.01$~T was applied and the time $t$-dependent magnetization $M(t)$ recorded.  Figure~\ref{Relaxation_compare} depicts the magnetic relaxation behavior of \ccia\ crystals with $x \geq 0.14$ measured with $H \parallel ab$ and $H \parallel c$,  where the time dependence of the relative change \mbox{$[M(t)-M(t = 0)]/M(0)$} is presented.  As seen from the figure, although little relaxation is apparent for $x = 0.14$, the crystals with $x = 0.17$, 0.25, and 0.35 show strong magnetization  relaxation for both $ab$-plane and $c$-axis magnetic fields. These results confirm metastable-state formation in the low-$T$ region for $0.17 \leq x \leq 0.35$. The composition $x = 0.14$ is thus close to the boundary between the \mbox{A-type} AFM phase and the FM cluster-glass phase.

We note that the parent compound \cca\ exhibits A-type AFM ordering with strong magnetic frustration within the $J_1-J_2$ Heisenberg model on a square lattice with a nearest-neighbor FM exchange interaction between the Co spins~\cite{Sapkota2017}. Increasing the Ir concentration results in an apparent increase in the FM interaction in these systems. The Ir substitution for Co occurs randomly leading to an increase of the randomly-distributed FM exchange interactions in the crystals. Thus it is plausible that in the presence of frustration and an increase in FM correlations induced by Ir~substitution, a low-$T$ magnetically-disordered glassy state is formed for $x \geq 0.17$ in \ccia\@. From Fig.~\ref{Relaxation_compare}, the relaxation is seen to be stronger for the magnetic field aligned in the  $ab$~plane compared to the $c$-axis alignment, suggesting that there is more in-plane magnetic disorder than between planes.

\begin{table}
\caption{\label{Tab.Relaxation} Parameters obtained from stretched exponential fits to the magnetic relaxation behaviour of \ccia\ compounds with $x =$ 0.17, 0.25, and 0.35.}
\begin{ruledtabular}
\begin{tabular}{|cccc|}	
Compound 									&		$H$ direction			& $\tau$ (sec)	 &  $\alpha$  \\
\hline\hline
Ca(Co$_{0.83}$Ir$_{0.17}$)$_{1.87}$As$_2$	& $H \parallel ab$	& 3598(107)	 &  0.44(1) \\
                                           	& $H \parallel c$	& 3269(108)	 &  0.45(1) \\
\hline
Ca(Co$_{0.75}$Ir$_{0.25}$)$_{1.89}$As$_2$	& $H \parallel ab$	& 3245(87)	 &  0.45(1) \\
                                           	& $H \parallel c$	& 2909(77)	 &  0.45(1) \\
\hline
Ca(Co$_{0.65}$Ir$_{0.35}$)$_{1.90}$As$_2$	& $H \parallel ab$	& 3840(116)	 &  0.46(1) \\
                                           	& $H \parallel c$	& 3189(106)	 &  0.42(1) \\
\end{tabular}
\end{ruledtabular}
\end{table}

In glassy systems, the relaxation of the magnetization~$M$ is often described by a stretched-exponential function with the time~$t$ dependence
\bea
\frac{M(t)}{M(t=0)}  =1- e^{-(t/\tau)^\alpha},
\label{Eq:Mrelax}
\eea
where $\alpha$ is the stretched-exponential exponent and $\tau$ is a characteristic relaxation time~\cite{Mydosh1993, Johnston2006}. A null value of $\alpha$ signifies no relaxation, whereas $\alpha = 1$ corresponds to a single magnetization relaxation time. Typically, magnetically-disordered glassy systems are characterized by a distribution of energy barriers where the value of $\alpha$ is between 0 and 1.

The time-dependent relaxation of the magnetization of the crystals with $x = 0.17$, 0.25, and 0.35 towards saturation is indeed well described by Eq.~(\ref{Eq:Mrelax}), as shown in Fig.~\ref{Relaxation_fit}.  The fitted parameters $\tau$ and~$\alpha$ for each~$x$ are listed in Table~\ref{Tab.Relaxation}. These results confirm the formation of a magnetically-disordered glassy state in the low-temperature region in \ccia\ crystals with $0.17\leq x \leq 0.35$.

\section{\label{Sec:Cp} Heat capacity}

\begin{figure*}
\includegraphics[width = 6in]{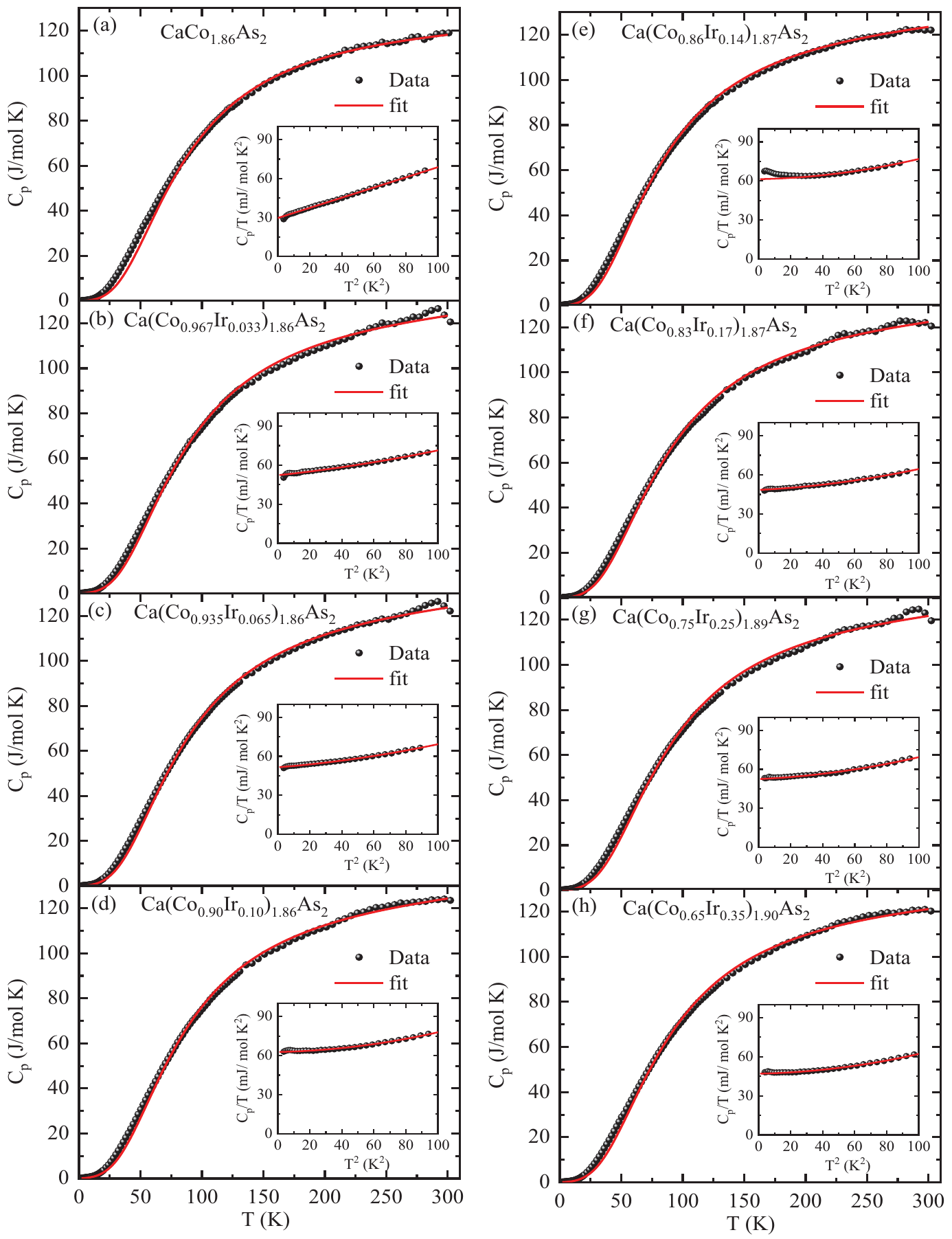}
\caption{Temperature dependence of the heat capacity C$_{\rm p}$ for \ccia\ crystals in zero field. The red solid lines are fits of the data by the Debye model using Eq.~(\ref{Eq:Debye_Fit}). Insets: C$_{\rm p}$/$T$ versus $T^2$ in the temperature range 1.8 K $\leq T \leq10$~K along with the fits by Eq.~(\ref{Eq.CpFit1}).}
\label{Heat_capacity_all}
\end{figure*}

Figure~\ref{Heat_capacity_all} shows the zero-field heat capacity $C_{\rm p}(T)$ of our \ccia\ crystals in the temperature range 1.8--300~K\@. The $C_{\rm p}(T)$ data saturate with increasing~$T$ to a value at 300~K close to the classical Dulong-Petit limit $C_{\rm p} = 3nR = 124.7$~J/mol\,K  where $R$ is the molar gas constant and here $n \approx 5$ is the number of atoms per formula unit.

\begin{table*}
\caption{\label{Tab.heatcapacityfit} The fitting parameters obtained from the analysis of heat capacity data. The listed parameters are the Sommerfeld coefficient ($\gamma$), the density of states at the Fermi energy ${\cal D}(E_{\rm F})$ derived from $\gamma$ using Eq.~(\ref{Eq:DfromGamma}), the lattice heat capacity coefficients $\beta$ and $\delta$ estimated from low-$T$ fit of C$_{\rm p}$/$T$ versus $T^2$ using Eq.~(\ref{Eq.CpFit1}), and Sommerfeld coefficient ($\gamma_{\rm D}$) along with the Debye temperature ($\Theta_{\rm D}$) determined by the fits of C$_{\rm p}$ versus~$T$ data using Eq.~(\ref{Eq:Debye_Fit}).}
\begin{ruledtabular}
\begin{tabular}{ccccccc}	
  											& $\gamma$	& ${\cal D}(E_{\rm F})$ & $\beta$ & $\delta$  &  $\gamma_{\rm D}$ 	& $\Theta_{\rm D}$ \\
Compound 									& (mJ mol$^{-1}$ K$^{-2}$)& (mJ mol$^{-1}$ K$^{-2}$) & (mJ mol$^{-1}$ K$^{-4}$) & ($\mu$J mol$^{-1}$ K$^{-6}$)  & (mJ mol$^{-1}$ K$^{-2}$) & (K)  \\
\hline
CaCo$_{1.86}$As$_2$                         				& 29.6(1) 	& 12.56(25) 	& 0.391(1)		& $\approx 0$	&	29.4(3)	& 	357(4)	\\
${\rm Ca(Co_{0.967}Ir_{0.033})_{1.86}As_2}$  			& 48.3(2) 	& 20.49(8) 	& 0.129(8)		& 0.61(6)		& 	34(2)		&	344(2)	\\
Ca(Co$_{0.935}$Ir$_{0.065}$)$_{1.86}$As$_2$    		&  51.6(1)	& 21.89(4)  	& 0.097(4) 	& 0.80(3)		&  	35(1)		&	344(2)	\\
Ca(Co$_{0.90}$Ir$_{0.10}$)$_{1.86}$As$_2$       	 	&  62.8(1)	& 26.64(4) 	& 0.025(6)		&  1.26(4)		& 	36(1)		&	337(2)	\\
Ca(Co$_{0.86}$Ir$_{0.14}$)$_{1.87}$As$_2$ 			& 61.4(4)	& 26.05(17)  	& 0.020(10)	&  1.32(7)		&  	33(1)		&	331(2)	\\		
Ca(Co$_{0.83}$Ir$_{0.17}$)$_{1.87}$As$_2$ 			& 54.4(1)	& 23.08(4)  	& 0.062(4) 	&  0.99(3)		& 	31(1)		&	349(2)	\\
Ca(Co$_{0.75}$Ir$_{0.25}$)$_{1.89}$As$_2$ 			& 52.5(5)	& 22.27(21)  	& 0.060(10)	& 1.11(7)		& 	30(2)		&	355(2)	\\
Ca(Co$_{0.65}$Ir$_{0.35}$)$_{1.90}$As$_2$ 			& 47.0(1)	& 19.94(4) 	& 0.066(5)		&  1.24(3)		& 	27(1)		&	349(2)	\\
\end{tabular}
\end{ruledtabular}
\end{table*}

The low-$T$ $C_{\rm p}(T)$ data in the temperature range 1.8~K~$\leq T \leq$~10~K for the \ccia\ crystals were analyzed using the relation
\bea
C_{\rm p}(T) = \gamma T+ \beta T^3 +\delta T^5,
\label{Eq.CpFit1}
\eea
where $\gamma$ is the Sommerfeld coefficient associated with the itinerant electrons and the last two terms constitute the low-$T$ lattice heat-capacity contribution. The insets in Figs.~\ref{Heat_capacity_all}(a)--\ref{Heat_capacity_all}(h) show  $C_{\rm p}(T)/T$ as a function of $T^2$ and the respective fits by Eq.~(\ref{Eq.CpFit1}).  The fitted values of $\gamma$, $\beta$, and $\delta$ are listed in Table~\ref{Tab.heatcapacityfit}. The $\gamma$ value increases significantly from 29.6(1) to 48.3(2) mJ mol$^{-1}$ K$^{-2}$ with only 3.3\% Ir substitution for Co in CaCo$_{1.86(2)}$As$_2$. This result reflects a sharp increase in the density of states at the Fermi energy ${\cal D}(E_{\rm F})$  with Ir substitution as determined from the relationship
\bea
{\cal D}_\gamma(E_{\rm F})\rm  {\left(  \frac{states}{eV\,f.u.} \right) = \frac{1}{2.357}\ \gamma \left(\frac{mJ}{mol\,K^2}\right)   },
\label{Eq:DfromGamma}
\eea
where this expression for ${\cal D}_\gamma(E_{\rm F})$ derived from~$\gamma$ includes the factor of two Zeeman degeneracy of the conduction carriers.  The values of ${\cal D}_\gamma(E_{\rm F})$ for the crystals are listed in Table~\ref{Tab.heatcapacityfit}.

\begin{figure}
\includegraphics[width = 3.4in]{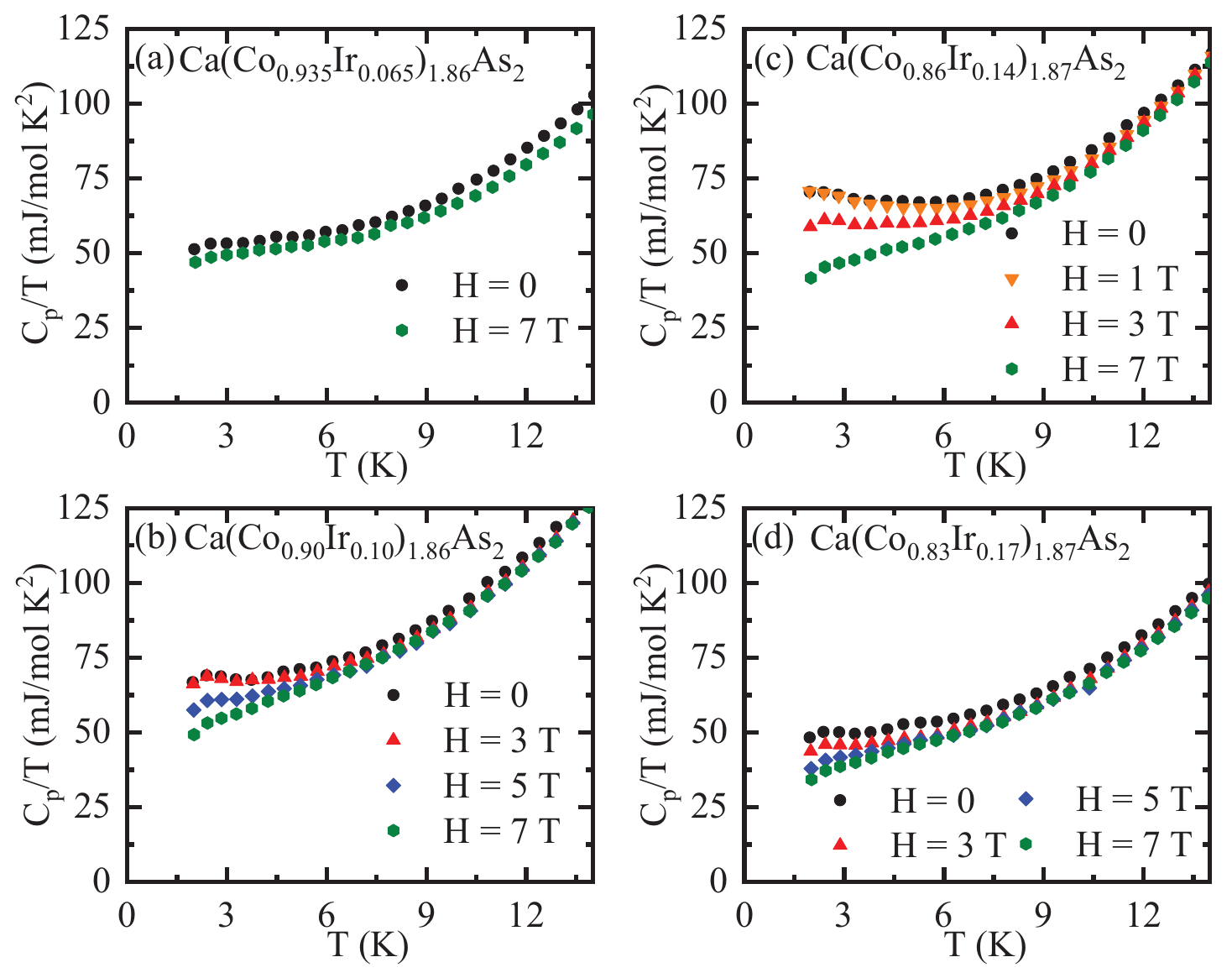}
\caption{C$_{\rm p}$/$T$ versus $T$ behavior at different applied magnetic fields in \ccia\ crystals for (a) $x = 0.065$, (b) $x =0.10$, (c) $x = 0.14$, and (d) $x = 0.17$. }
\label{Heat_capacity_CpbyT_T}
\end{figure}

Low-$T$ upturns are observed in the C$_{\rm p}$/$T$ versus~$T$ plots in Fig.~\ref{Heat_capacity_CpbyT_T} for the crystals with $x = 0.10$ and~0.14.  The upturn is more pronounced for the latter composition.  These upturns are not fitted by  Eq.~(\ref{Eq.CpFit1}) down to the lowest measured temperature and will therefore be refitted below.

\begin{figure}
\includegraphics[width = 3.3in]{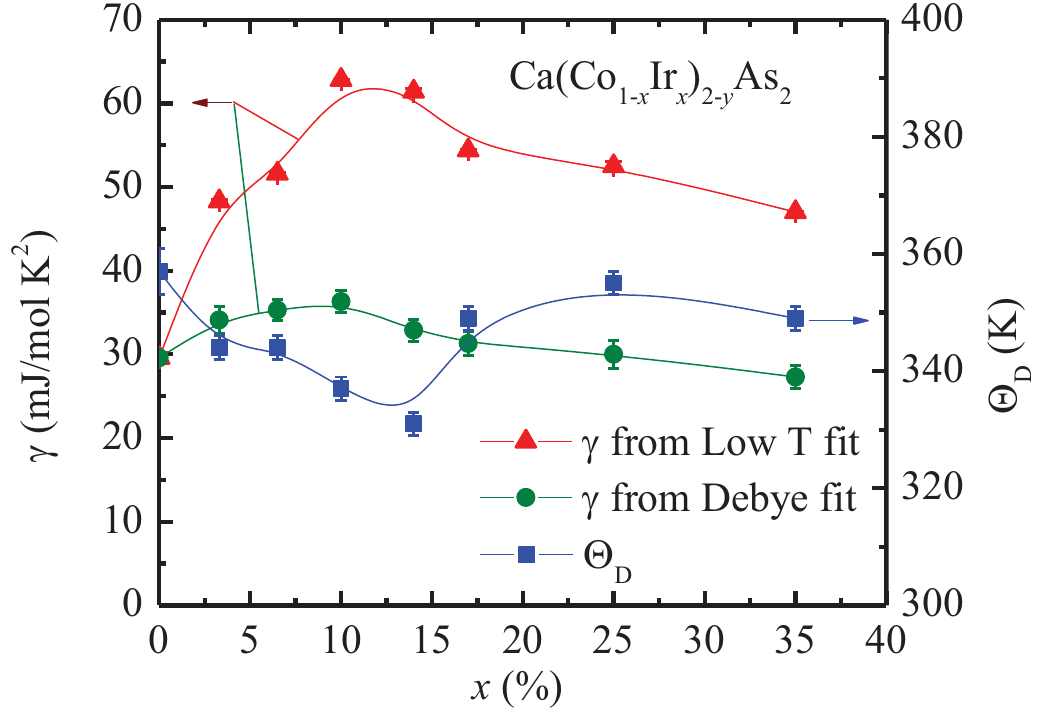}
\caption{Sommerfeld coefficient $\gamma$ from the low-$T$ heat capacity fit by Eq.~(\ref{Eq.CpFit1}),  and $\gamma_{\rm D}$ and $\Theta_{\rm D}$ from the Debye fit in Eq.~\ref{Eq:Debye_Fit} for the \ccia\ crystals.}
\label{Heat_capacity_parameters}
\end{figure}

The  $C_{\rm p}(T)$ data were analyzed over the entire temperature range of the measurements using the relation
\bea
C_{\rm p}(T) &=& \gamma_{\rm D} T+ nC_{\rm V\,Debye}(T),\label{Eq:Debye_Fit} \\*
C_{\rm V\,Debye}(T) &=& 9R \left(\frac{T}{\Theta_{\rm D}}\right)^3\int_{0}^{\Theta_{\rm D}/T}\frac{x^4e^x}{(e^x-1)^2} dx,\nonumber
\eea
where $n$ is the number of atoms per formula unit,  $\gamma_{\rm D}$ is the Sommerfeld coefficient derived from the present fit, $C_{\rm V\,Debye}$ is the Debye lattice heat capacity per mole of atoms at constant volume, and $\Theta_{\rm D}$ is the Debye temperature. The fitted values of $\gamma_{\rm D}$  and $\Theta_{\rm D}$ for all the crystals are listed in Table~\ref{Tab.heatcapacityfit}. The values of  $\gamma$ and $\gamma_{\rm D}$ are the same for the parent compound with $x=0$ but differ significantly from each other for the Ir-substituted crystals with $x>0$ as shown in Table~\ref{Tab.heatcapacityfit}  and Fig.~\ref{Heat_capacity_parameters}, where $\gamma_{\rm D}\sim\gamma/2$.

\begin{table*}[ht!]
\caption{\label{Tab.Quantumcritical} Parameters obtained from fitting the $C_{\rm p}$/$T$ versus $T$ behavior by Eq.~(\ref{Eq.Cp_SF_Fit}) in the temperature range 1.8--10 K for $x = 0.10$ and 0.14.}
\begin{ruledtabular}
\begin{tabular}{cccccc}	
  											& $\gamma_{\rm {SF}}$	& $\beta_{\rm {SF}}$ & $\delta_{\rm {SF}}$  &  $\kappa$ 	& $T_{\rm {SF}}$ \\
Compound 									& (mJ mol$^{-1}$ K$^{-2}$)& (mJ mol$^{-1}$ K$^{-4}$) & ($\mu$J mol$^{-1}$ K$^{-6}$)  & (mJ mol$^{-1}$ K$^{-2}$) & (K)  \\
\hline
Ca(Co$_{0.90}$Ir$_{0.10}$)$_{1.86}$As$_2$       	 &  57.6(2)		 & 	0.06(3)	&  	1.3(3)		& 	-3.1(8)		&	18(3)	\\
Ca(Co$_{0.86}$Ir$_{0.14}$)$_{1.87}$As$_2$ 		& 	53.2(7)	&  		0.06(1)	&  1.5(1)	&  		-7.6(4)		&	14(2)	\\		
\end{tabular}
\end{ruledtabular}
\end{table*}

\begin{figure}
\includegraphics[width = 2.7in]{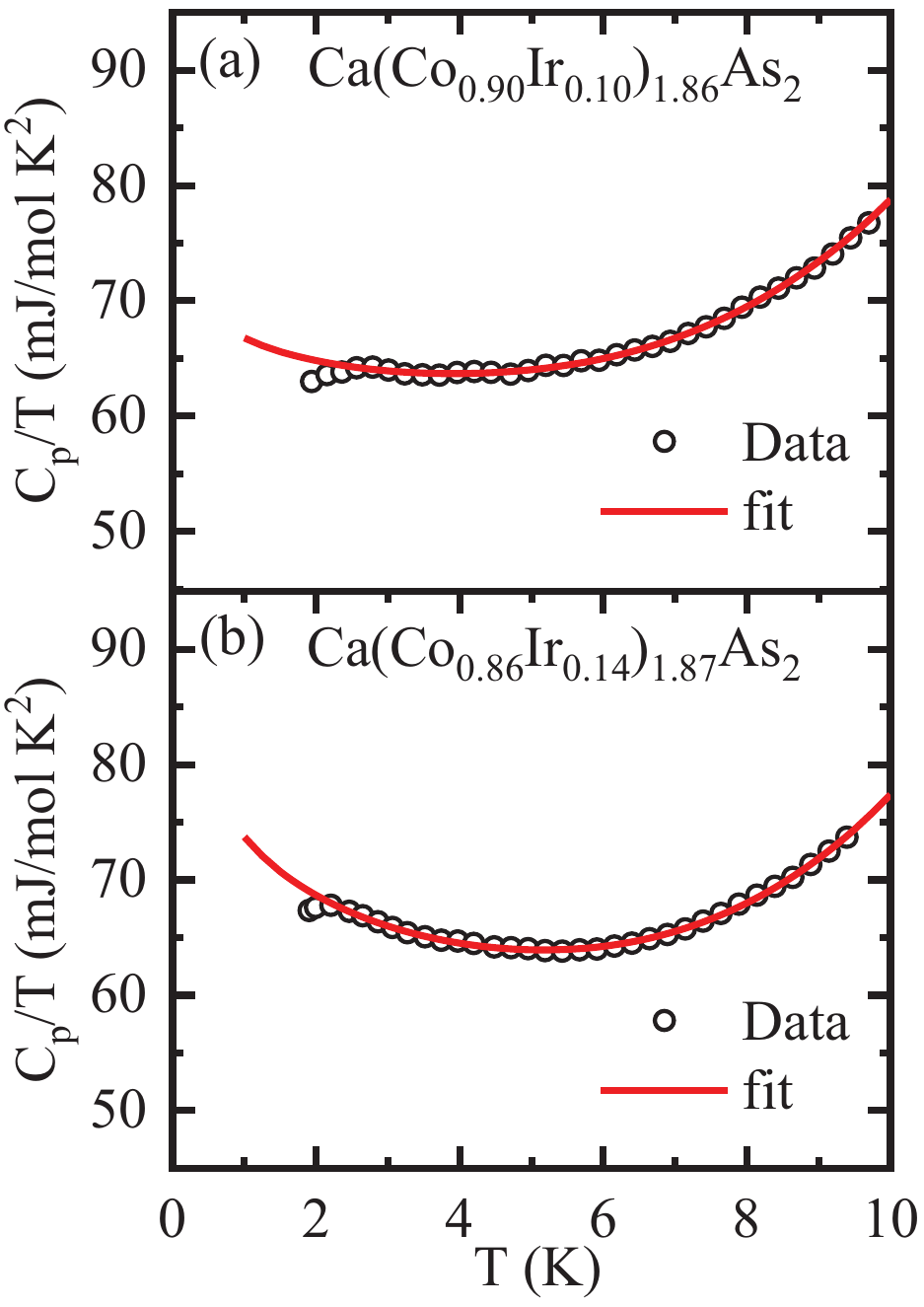}
\caption{Zero-field C$_{\rm p}$/$T$ versus $T$ data of \ccia\ crystals with (a) $x =$ 0.10 and (b)~$x =$ 0.14, in the temperature range 1.8 to 10 K\@. The solid red curves are fits by Eq.~(\ref{Eq.Cp_SF_Fit}).}
\label{Heat_capacity_QCP}
\end{figure}

In order to investigate the origin of the low-$T$ upturns in $C_{\rm p}(T)/T$ versus~$T$ for the \ccia\ crystals with $x = 0.10$ and 0.14 in Fig.~\ref{Heat_capacity_CpbyT_T}, the dependence of $C_{\rm p}(T)$ on applied $c$-axis fields from 0 to~7~T was measured for  $x=0.10$ and~0.14 together with analogous measurements of the neighboring compositions $x =0.065$ and 0.17.   Figure~\ref{Heat_capacity_CpbyT_T} shows that the magnetic field strongly alters C$_{\rm p}$/$T$ versus~$T$ for $x = 0.14$, moderately alters those of the crystals with $x = 0.10$ and~0.17, and has almost no influence for $x = 0.065$. The low-$T$ upturn for $x = 0.14$ is strongly suppressed by the applied field. Such types of behavior have been previously observed for different compounds with FM quantum-critical fluctuations~\cite{WuPNAS2014, Nicklas1999} and were also recently reported to occur in isostructural \scna\ crystals~\cite{Sangeetha2019scna}.

In this context we recall that with increasing Ir substitution for Co in CaCo$_{1.86(2)}$As$_2$, FM interactions increase  significantly and FM clustering occurs for $x \geq 0.17$. Furthermore, the above analyses revealed that the $x = 0.14$ composition is close to the phase boundary between the competing A-type AFM phase and magnetically disordered FM cluster-glass phase. Thus, it seems reasonable to interpret the upturn in $C/T$ versus $T$ at low $T$ for $x = 0.14$ as arising from FM quantum-critical fluctuations associated with a quantum-critical composition between the AFM and FM cluster-glass states.

\begin{figure}
\includegraphics[width = 3.4in]{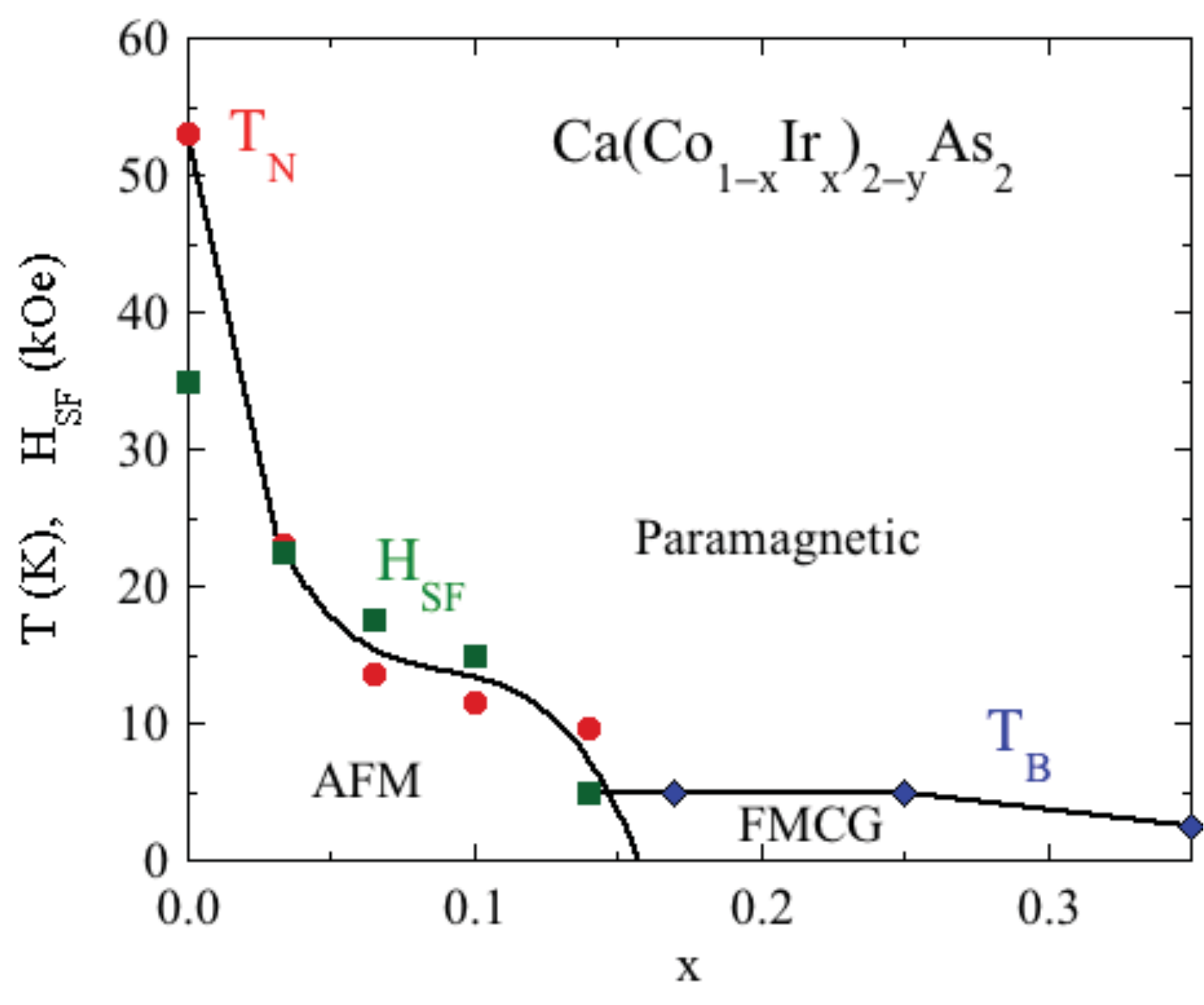}
\caption{Approximate phase diagram of the  \ccia\ system.  The N\'eel temperatures $T_{\rm N}$ (filled red circles), spin flop fields~$H_{\rm SF}$ (filled green squares), and blocking temperatures $T_{\rm B}$ (filled blue diamonds) from Table~\ref{Tab.chidata} are plotted versus the Ir concentration~$x$.  The estimated phase boundaries between the A-type antiferromagntic (AFM), paramagnetic, and ferromagnetic cluster-glass (FMCG) phases are indicated by black lines. }
\label{Phase_diagram}
\end{figure}

In some systems a $C/T \sim \ln T$ contribution attributed to quantum-critical fluctuations has been included to fit the low-$T$ $C_{\rm p}(T)$ data in addition to electronic and lattice heat capacity contributions~\cite{Sangeetha2019scna, Flude1968, Millis1993}.  Therefore, for our crystals with $x = 0.10$ and 0.14, we analyzed the $C_{\rm p}/T$ versus $T$ data using the relation
\bea
\frac{C_{\rm p}(T)}{T} = \gamma_{\rm SF} + \beta_{\rm SF}T^2 +\delta_{\rm SF} T^4 + \kappa{\rm ln}(T/T_{\rm SF}),
\label{Eq.Cp_SF_Fit}
\eea
where $\gamma_{\rm SF}$ is the Sommerfeld coefficient determined from this fit, $\beta_{\rm SF}$ and $\delta_{\rm SF}$ are the lattice heat capacity coefficients from this fit, $\kappa$ is the spin fluctuation coefficient, and $T_{\rm SF}$ is the spin fluctuation temperature. The low-$T$ upturns seen in the data for the $x = 0.10$ and 0.14 crystals was well fitted by incorporating the latter term, as shown in Fig.~\ref{Heat_capacity_QCP}. The fitted parameters are listed in Table.~\ref{Tab.Quantumcritical}.

The upturn in $C_{\rm p}/T$ versus~$T$ for the $x =0.14$ crystal is most pronounced and the analysis suggests that this composition exhibits FM quantum-critical fluctuations, but where a FM quantum critical point (QCP) is preempted by the occurrence of a QCP associated with the AFM transition. Such avoided FM QCPs have been observed in various other magnetic systems~\cite{Lausberg2012, Lengyel2015, Hu2015}.

An approximate  phase diagram in the temperature-composition plane of the \ccia\ system is shown in Fig.~\ref{Phase_diagram} based on the presently-available $T_{\rm N}(x)$ and $T_{\rm B}(x)$ data.  Also included is the concentration dependence of the spin-flop field $H_{\rm SF}$ which is expected to be correlated with $T_{\rm N}$ and is therefore used to help delineate the boundaries between the AFM, PM, and FMCG phases.


\section{\label{ConcRem} Summary}

In this work, we investigated the physical properties of Ir-substituted \ccia\ single crystals with $0\leq x \leq 0.35$ and $0.10\leq y \leq 0.14$ that were grown out of Co-Ir-As self flux. Room-temperature x-ray diffraction measurements showed that all the crystals form in the collapsed-tetragonal structure, as does the parent compound \cca. The SEM-EDS results showed that the vacancy concentration on the Co site is sensitive to the Ir concentration~$x$, which changes from 7\% in the parent \cca\ to 5\% in the 35\% Ir-substituted composition. Both the tetragonal lattice parameters $a =b$ and $c$ were found to increase nonlinearly with ~$x$, yielding a nonlinear unit cell volume versus~$x$.

The temperature dependence of the static magnetic susceptibility $\chi$ shows that the A-type AFM transition temperature $T_{\rm N} = 52$~K of \cca\ rapidly drops to 23~K with only 3.3\% Ir substitution. Increasing $x$ dramatically enhances ferromagnetic fluctuations, both in-plane and out-of-plane, and further decreases $T_{\rm N}$ which approaches zero for $x = 0.14$. A phase transition then occurs at $x\approx 0.16$ to a ferromagnetic cluster-glass phase for $0.17 \leq x \leq 0.35$ below a blocking temperature $T_{\rm B} \lesssim 5$~K, exemplified by observations of magnetization irreversibility between zero-field-cooling and field-cooling measurements, a small spontaneous ferromagnetic component to the ordering, and spin-glass-type stretched-exponential dynamics in the time dependence of the magnetization response to a small applied field.

The zero-field heat capacity $C_{\rm p}(T)$ reveals a strong increase in the Sommerfeld heat-capacity coefficient $\gamma$ with Ir substitution, signifying a corresponding increase in the electronic density of states at the Fermi energy in the Ir-substituted \ccia\ crystals. A low-temperature upturn in the $C_{\rm p}/T$ versus $T$ is observed for $x = 0.10$ and 0.14, with the strongest such contribution for $x=0.14$, near the boundary between the A-type AFM phase and the FMCG phase. The upturn in $C_{\rm p}/T$ versus~$T$ for $x=0.14$ is fitted well by a logarithmic temperature dependence that we associate with ferromagnetic quantum-critical spin fluctuations. The observed upturn in  $C_{\rm p}/T$ versus $T$ is strongly suppressed by an $H = 7$~T magnetic field, consistent with this interpretation.

Thus increasing the random Ir-substitution for Co in \cca\ results in a strong increase in FM interactions that results in a phase transition at $x\approx 0.16$ from A-type AFM to a FMCG phase   The composition $x = 0.14$ which is close the transition composition exhibits a signature of ferromagnetic quantum fluctuations.   We infer that the FMCG phase at larger $x$ ($x =  0.17$, 0.25, 0.35) arises from a competition between AFM and FM interactions in conjunction with crystallographic disorder accompanying Ir substitution.


\acknowledgments

This research was supported by the U.S. Department of Energy, Office of Basic Energy Sciences, Division of Materials Sciences and Engineering.  Ames Laboratory is operated for the U.S. Department of Energy by Iowa State University under Contract No.~DE-AC02-07CH11358.


\begin{thebibliography}{99}

\bibitem{Rotter2008} M. Rotter, M. Tegel, and D. Johrendt, Superconductivity at 38 K in the Iron Arsenide (Ba$_{1-x}$K$_x$)Fe$_2$As$_2$, Phys. Rev. Lett. {\bf 101}, 107006 (2008).

\bibitem{Johnston2010} D. C. Johnston,  The puzzle of high temperature superconductivity in layered iron pnictides and chalcogenides, Adv. Phys. {\bf 59}, 803 (2010).

\bibitem{Canfield2010} P. C. Canfield and S. L. Bud'ko,  Magnetism and its microscopic origin in iron-based high-temperature superconductors, Annu. Rev. Condens. Matter Phys. {\bf 1}, 27 (2010).

\bibitem{Paglione2010} J. Paglione and R. L. Greene, High-temperature superconductivity in iron-based materials, Nat. Phys. {\bf 6}, 645 (2010).

\bibitem{Fernandes2010} R. M. Fernandes, D. K. Pratt, W. Tian, J. Zarestky, A. Kreyssig, S. Nandi, M. G. Kim, A. Thaler, N. Ni, P. C. Canfield, R. J. McQueeney, J. Schmalian, and A. I. Goldman, Unconventional pairing in the iron arsenide superconductors, Phys. Rev. B {\bf 81}, 140501(R) (2010).

\bibitem{Stewart2011} G. R. Stewart,  Superconductivity in iron compounds, Rev. Mod. Phys. {\bf 83}, 1589 (2011).

\bibitem{Scalapino2012} D. J. Scalapino, A common thread: The pairing interaction for unconventional superconductors, Rev. Mod. Phys. {\bf 84}, 1383 (2012).

\bibitem{Dai2012} P. Dai, J. Hu, and E. Dagotto,  Magnetism and its microscopic origin in iron-based high-temperature superconductors, Nat. Phys. {\bf 8}, 709 (2012).

\bibitem{Dagotto2013} E. Dagotto, The unexpected properties of alkali metal iron selenide superconductors, Rev. Mod. Phys. {\bf 85}, 849 (2013).

\bibitem{Dai2015} P. Dai, Antiferromagnetic order and spin dynamics in iron-based superconductors, Rev. Mod. Phys. {\bf 87}, 855 (2015).

\bibitem{Anand2012} For a review of collapsed- and uncollapsed-tetragonal materials, see Sec.~VIII of V. K. Anand, P. K. Perera, A. Pandey, R. J. Goetsch, A. Kreyssig, and D. C. Johnston, Crystal growth and physical properties of ${\rm SrCu_2As_2}$, ${\rm SrCu_2Sb_2}$ and ${\rm BaCu_2Sb_2}$, Phys. Rev. B {\bf 85}, 174509 (2012).

\bibitem{Sasmal2008} K. Sasmal, B. Lv, B. Lorenz, A. M. Guloy, F. Chen, Y.-Y. Xue, and C.-W. Chu, Superconducting Fe-Based Compounds ($A_{1-x}$Sr$_x$)Fe$_2$As$_2$ with $A$ = K and Cs with Transition Temperatures up to 37~K, Phys. Rev. Lett. {\bf 101},  107007 (2008).

\bibitem{Jasper2008} A. Leithe-Jasper, W. Schnelle, C. Geibel, and H. Rosner, Superconducting State in SrFe$_{2 - x}$Co$_x$As$_2$ Internal Doping of the Iron Arsenide Layers, Phys. Rev. Lett. {\bf 101}, 207004 (2008).

\bibitem{Kumar2009} N. Kumar, S. Chi, Y. Chen, K. G. Rana, A. K. Nigam, A. Thamizhavel, W. Ratcliff, II, S. K. Dhar, and J. W. Lynn, Evolution of the bulk properties, structure, magnetic order, and superconductivity with Ni doping in CaFe$_{2 - x}$Ni$_x$As$_2$, Phys. Rev. B {\bf 80}, 144524 (2009).

\bibitem{Ren2009} Z. Ren, Q. Tao, S. Jiang, C. Feng, C. Wang, J. Dai, G. Cao, and Z. Xu, Superconductivity Induced by Phosphorus Doping and Its Coexistence with Ferromagnetism in EuFe$_2$(As$_{0.7}$P$_{0.3}$)$_2$, Phys. Rev. Lett. {\bf 102}, 137002 (2009).

\bibitem{Rotter2008Angew} M. Rotter, M. Pangerl, M. Tegel, and D. Johrendt, Superconductivity and Crystal Structures of (Ba$_{1-x}$K$_x$)Fe$_2$As$_2$ ($x = 0-1$), Angew. Chem., Int. Ed. {\bf 47}, 7949 (2008).

\bibitem{Bukowski2010} Z. Bukowski, S. Weyeneth, R. Puzniak, J. Karpinski, and B. Batlogg, Bulk superconductivity at 2.6~K in undoped RbFe$_2$As$_2$, Physica C {\bf 470}, S328 (2010).

\bibitem{Sefat2009} A. S. Sefat, D. J. Singh, R. Jin, M. A. McGuire, B. C. Sales, and D. Mandrus, Renormalized behavior and proximity of \bca\ to a magnetic quantum critical point, Phys. Rev. B {\bf 79}, 024512 (2009).

\bibitem{Anand2014} V. K. Anand, D. G. Quirinale, Y. Lee, B. N. Harmon, Y. Furukawa, V. V. Ogloblichev, A. Huq, D. L. Abernathy, P. W. Stephens, R. J. McQueeney, A. Kreyssig, A. I. Goldman, and D. C. Johnston, Crystallography and physical properties of \bca\, Ba$_{0.94}$K$_{0.06}$Co$_2$As$_2$, and Ba$_{0.78}$K$_{0.22}$Co$_2$As$_2$, Phys. Rev. B {\bf 90}, 064517 (2014).

\bibitem{Pandey2013} A. Pandey, D. G. Quirinale, W. Jayasekara, A. Sapkota, M. G. Kim, R. S. Dhaka, Y. Lee, T. W. Heitmann, P. W. Stephens, V. Ogloblichev, A. Kreyssig, R. J. McQueeney, A. I. Goldman, A. Kaminski, B. N. Harmon, Y. Furukawa, and D. C. Johnston, Crystallographic, electronic, thermal, and magnetic properties of single-crystal \sca, Phys. Rev. B {\bf 88}, 014526 (2013).

\bibitem{Quirinale2013} D. G. Quirinale, V. K. Anand, M. G. Kim, A. Pandey, A. Huq, P. W. Stephens, T. W. Heitmann, A. Kreyssig, R. J. McQueeney, D. C. Johnston, and A. I. Goldman, Crystal and magnetic structure of CaCo$_{1.86}$As$_2$ studied by x-ray and neutron diffraction, Phys. Rev. B {\bf 88}, 174420 (2013).

\bibitem{Anand2014Ca} V. K. Anand, R. S. Dhaka, Y. Lee, B. N. Harmon, A. Kaminski, and D. C. Johnston, Physical properties of metallic antiferromagnetic CaCo$_{1.86}$As$_2$ single crystals, Phys. Rev. B {\bf 89}, 214409 (2014).

\bibitem{Bing2019sca} B. Li, B. G. Ueland, W. T. Jayasekara, D. L. Abernathy, N. S. Sangeetha, D. C. Johnston, Q.-P. Ding, Y. Furukawa, P. P. Orth, A. Kreyssig, A. I. Goldman, and R. J. McQueeney, Competing magnetic phases and itinerant magnetic frustration in \sca, Phys. Rev. B {\bf 100}, 054411 (2019).

\bibitem{Jayasekara2013} W. Jayasekara, Y. Lee, A. Pandey, G. S. Tucker, A. Sapkota, J. Lamsal, S. Calder, D. L. Abernathy, J. L. Niedziela, B. N. Harmon, A. Kreyssig, D. Vaknin, D. C. Johnston, A. I. Goldman, and R. J. McQueeney, Stripe Antiferromagnetic Spin Fluctuations in \sca, Phys. Rev. Lett. {\bf 111}, 157001 (2013).

\bibitem{Wiecki2015} P. Wiecki, V. Ogloblichev, A. Pandey, D. C. Johnston, and Y. Furukawa, Coexistence of antiferromagnetic and ferromagnetic spin correlations in \sca\ revealed by $^{59}$Co and $^{75}$As NMR, Phys. Rev. B {\bf 91}, 220406(R) (2015).

\bibitem{Li2019} Y. Li, Z. Yin, Z. Liu, W. Wang, Z. Xu, Y. Song, L. Tian, Y. Huang, D. Shen, D. L. Abernathy, J. L. Niedziela, R. A. Ewings, T. G. Perring, D. M. Pajerowski, M. Matsuda, P. Bourges, E. Mechthild, Y. Su, and P. Dai, Coexistence of Ferromagnetic and Stripe Antiferromagnetic Spin Fluctuations in \sca, Phys. Rev. Lett. {\bf 122}, 117204 (2019).

\bibitem{Wiecki2015b} P. Wiecki, B. Roy, D. C. Johnston, S. L. Bud'ko, P. C. Canfield, and Y. Furukawa, Competing Magnetic Fluctuations in Iron Pnictide Superconductors: Role of Ferromagnetic Spin Correlations Revealed by NMR, Phys. Rev. Lett {\bf 115}, 137001 (2015).

\bibitem{Cheng2012} B. Cheng, B. F. Hu, R. H. Yuan, T. Dong, A. F. Fang, Z. G. Chen, G. Xu, Y. G. Shi, P. Zheng, J. L. Luo, and N. L. Wang, Field induced spin flop transitions in single-crystalline \cca, Phys. Rev. B {\bf 85}, 144426 (2012).

\bibitem{Ying2012} J. J. Ying, Y. J. Yan, A. F. Wang, Z. J. Xiang, P. Cheng, G. J. Ye, and X. H. Chen, Metamagnetic transition in Ca$_{1-x}$Sr$_x$Co$_2$As$_2$ ($x$ = 0 and 0.1) single crystals, Phys. Rev. B {\bf 85}, 214414 (2012).

\bibitem{Sapkota2017} A. Sapkota, B. G. Ueland, V. K. Anand, N. S. Sangeetha, D. L. Abernathy, M. B. Stone, J. L. Niedziela, D. C. Johnston, A. Kreyssig, A. I. Goldman, and R. J. McQueeney, Effective One-Dimensional Coupling in the Highly Frustrated Square-Lattice Itinerant Magnet CaCo$_{2-y}$As$_2$, Phys. Rev. Lett. {\bf 119}, 147201 (2017).

\bibitem{Ying2013} J. J. Ying, J. C. Liang, X. G. Luo, Y. J. Yan, A. F. Wang, P. Cheng, G. J. Ye, J. Q. Ma, and X. H. Chen, The magnetic phase diagram of \csca\ single crystals, Europhys. Lett. {\bf 104}, 67005 (2013).

\bibitem{Sangeetha2017} N. S. Sangeetha, V. Smetana, A.-V. Mudring, and D. C. Johnston, Anomalous Composition-Induced Crossover in the Magnetic Properties of the Itinerant-Electron Antiferromagnet \csca\, Phys. Rev. Lett. {\bf 119}, 257203 (2017).

\bibitem{Bing2019csca} B. Li, Y. Sizyuk, N. S. Sangeetha, J. M. Wilde, P. Das, W. Tian, D. C. Johnston, A. I. Goldman, A. Kreyssig, P. P. Orth, R. J. McQueeney, and B. G. Ueland, Antiferromagnetic stacking of ferromagnetic layers and doping-controlled phase competition in \csca, Phys. Rev. B {\bf 100}, 024415 (2019).

\bibitem{Sangeetha2019scna} N. S. Sangeetha, L.-L. Wang, A. V. Smirnov, V. Smetana, A.-V. Mudring, D. D. Johnson, M. A. Tanatar, R. Prozorov, and D. C. Johnston, Non-Fermi-liquid types of behavior associated with a magnetic quantum critical point in \scna\ single crystals, Phys. Rev. B {\bf 100}, 094447 (2019).

\bibitem{YLi2019} Y. Li, Z. Liu, Z. Xu, Y. Song, Y. Huang, D. Shen, N. Ma, A. Li, S. Chi, M. Frontzek, H. Cao, Q. Huang, W. Wang, Y. Xie, R. Zhang, Y. Rong, W. A. Shelton, D. P. Young, J. F. DiTusa, and P. Dai, Flat-band magnetism and helical magnetic order in Ni-doped \sca, Phys. Rev. B {\bf 100}, 094446 (2019).

\bibitem{Wilde2019} J. M. Wilde, A. Kreyssig, D. Vaknin, N. S. Sangeetha, B. Li, W. Tian, P. P. Orth, D. C. Johnston, B. G. Ueland, and R. J. McQueeney, Helical magnetic ordering in \scna, Phys. Rev. B {\bf 100}, 161113(R)(2019).

\bibitem{Shen2018} S. Shen, S. Feng, Z. Lin, Z. Wang, and W. Zhong, Ferromagnetic behavior induced by La-doping in \sca, J. Mater. Chem. C {\bf 6}, 8076 (2018).

\bibitem{Shen2019} S. Shen, W. Zhong, D. Li, Z. Lin, Z. Wang, X. Gu, and S. Feng, Itinerant ferromagnetism induced by electron doping in \sca, Inorg. Chem. Commun. {\bf 103}, 25 (2019).

\bibitem{Jayasekara2017} W. T. Jayasekara, A. Pandey, A. Kreyssig, N. S. Sangeetha, A. Sapkota, K. Kothapalli, V. K. Anand, W. Tian, D. Vaknin, D. C. Johnston, R. J. McQueeney, A. I. Goldman, and B. G. Ueland, Suppression of magnetic order in CaCo$_{1.86}$As$_2$ with Fe substitution: Magnetization, neutron diffraction, and x-ray diffraction studies of Ca(Co$_{1-x}$Fe$_x$)$_y$As$_2$, Phys. Rev. B {\bf 95}, 064425 (2017).

\bibitem{Smidman2017} M. Smidman, M. B. Salamon, H. Q. Yuan, and D. F. Agterberg, Superconductivity and spin-orbit coupling in non-centrosymmetric materials: a review, Rep. Prog. Phys. {\bf 80}, 036501 (2017).

\bibitem{Marco2010} M. A. Laguna-Marco, D. Haskel, N. Souza-Neto, J. C. Lang, V. V. Krishnamurthy, S. Chikara, G. Cao, and M. van Veenendaal, Orbital Magnetism and Spin-Orbit Effects in the Electronic Structure of BaIrO$_3$, Phys. Rev. Lett. {\bf 105}, 216407 (2010).

\bibitem{Ng2000} K. K. Ng and M. Sigrist, The role of spin-orbit coupling for the superconducting state in Sr$_2$RuO$_4$, Europhys. Lett. {\bf 49}, 473 (2000).

\bibitem{Rhodes2015} D. Rhodes, S. Das, Q. R. Zhang, B. Zeng, N. R. Pradhan, N. Kikugawa, E. Manousakis, and L. Balicas, Role of spin-orbit coupling and evolution of the electronic structure of WTe$_2$ under an external magnetic field, Phys. Rev. B {\bf 92}, 125152 (2015).

\bibitem{Plumb2014} K. W. Plumb, J. P. Clancy, L. J. Sandilands, V. V. Shankar, Y. F. Hu, K. S. Burch, H.-Y. Kee, and Y.-J. Kim, $\alpha$-RuCl$_3$: A spin-orbit assisted Mott insulator on a honeycomb lattice, Phys. Rev. B {\bf 90}, 041112(R) (2014).

\bibitem{Ma2017} M. Ma, P. Bourges, Y. Sidis, Y. Xu, S. Li, B. Hu, J. Li, F. Wang, and Y. Li, Prominent Role of Spin-Orbit Coupling in FeSe Revealed by Inelastic Neutron Scattering, Phys. Rev. X {\bf 7}, 021025 (2017).

\bibitem{APEX2015} APEX3, Bruker AXS Inc., Madison, Wisconsin, USA, 2015.

\bibitem{SAINT2015} SAINT, Bruker AXS Inc., Madison, Wisconsin, USA, 2015.

\bibitem{Krause2015} L. Krause, R. Herbst-Irmer, G. M. Sheldrick, and D. J. Stalke, Appl. Crystallogr. {\bf 48}, 3 (2015).

\bibitem{Sheldrick2015A} G. M. Sheldrick, SHELTX -- Integrated space-group and crystal-structure determination, Acta Crystallogr. A {\bf 71}, 3 (2015).

\bibitem{Sheldrick2015C} G. M. Sheldrick, Crystal structure refinement with SHELXL.  Acta Crystallogr.~C {\bf 71}, 3 (2015).

\bibitem{Fisher1962} M. E. Fisher, Relation between the Specific Heat and Susceptibility of an Antiferromagnet, Philos. Mag. {\bf 7}, 1731 (1962).

\bibitem{Pakhira2018} S. Pakhira, C. Mazumdar, R. Ranganathan, and S. Giri, Magnetic phase inhomogeneity in frustrated intermetallic compound Sm$_2$Ni$_{0.87}$Si$_{2.87}$, J. Alloys Compd. {\bf 742}, 391 (2018).

\bibitem{DXLi2003} D. X. Li, S. Nimori, Y. Shiokawa, Y. Haga, E. Yamamoto, and Y. Onuki, Ferromagnetic cluster glass behavior in U$_2$IrSi$_3$, Phys. Rev. B {\bf 68}, 172405 (2003).

\bibitem{Freitas2001} R. S. Freitas, L. Ghivelder, F. Damay, F. Dias, and L. F. Cohen, Magnetic relaxation phenomena and cluster glass properties of La$_{0.7 - x}$Y$_x$Ca$_{0.3}$MnO$_3$ manganites, Phys. Rev. B {\bf 64}, 144404 (2001).

\bibitem{Nam1999} D. N. H. Nam, K. Jonason, P. Nordblad, N. V. Khiem, and N. X. Phuc, Coexistence of ferromagnetic and glassy behavior in the La$_{0.5}$Sr$_{0.5}$CoO$_3$ perovskite compound, Phys. Rev. B {\bf 59}, 4189 (1999).

\bibitem{Anand2012PrRhSn3} V. K. Anand, D. T. Adroja, and A. D. Hillier, Ferromagnetic cluster spin-glass behavior in PrRhSn$_3$, Phys. Rev. B {\bf 85}, 014418 (2012).

\bibitem{Mydosh1993} J. A. Mydosh, {\it Spin Glasses: An Experimental Introduction} (Taylor \& Francis, London, 1993).

\bibitem{Binder1986} K. Binder and A. P. Young, Spin glasses: Experimental facts, theoretical concepts, and open questions, Rev. Mod. Phys. {\bf 58}, 801 (1986).

\bibitem{Nordblad1986} P. Nordblad, P. Svedlindh, L. Lundgren, and L. Sandlund, Time decay of the remanent magnetization in a Cu$Mn$ spin glass, Phys. Rev. B {\bf 33}, 645(R) (1986).

\bibitem{Chu1994} D. Chu, G. G. Kenning, and R. Orbach, Dynamic measurements in a Heisenberg spin glass: CuMn, Phys. Rev. Lett. {\bf 72}, 3270 (1994).

\bibitem{Johnston2006} D. C. Johnston, Stretched Exponential Relaxation Arising from a Continuous Sum of Exponential Decays, Phys. Rev. B {\bf 74}, 184430 (2006).

\bibitem{WuPNAS2014} L. S. Wu, M. S. Kim, K. Park, A. M. Tsvelik, and M. C. Aronson, Quantum critical fluctuations in layered YFe$_2$Al$_{10}$, Proc. Natl. Acad. Sci. U.S.A. {\bf 111}, 14088 (2014).

\bibitem{Nicklas1999} M. Nicklas, M. Brando, G. Knebel, F. Mayr, W. Trinkl, and A. Loidl, Non-Fermi-Liquid Behavior at a Ferromagnetic Quantum Critical Point in Ni$_x$Pd$_{1-x}$, Phys. Rev. Lett. {\bf 82}, 4268 (1999).

\bibitem{Flude1968} P. Fulde and A. Luther, Effects of Impurities on Spin Fluctuations in Almost Ferromagnetic Metals, Phys. Rev. {\bf 170}, 570 (1968).

\bibitem{Millis1993} A. J. Millis, Effect of a nonzero temperature on quantum critical points in itinerant fermion systems, Phys. Rev. B {\bf 48}, 7183 (1993).

\bibitem{Lausberg2012} S. Lausberg, J. Spehling, A. Steppke, A. Jesche, H. Luetkens, A. Amato, C. Baines, C. Krellner, M. Brando, C. Geibel, H.-H. Klauss, and F. Steglich, Avoided Ferromagnetic Quantum Critical Point: Unusual Short-Range Ordered State in CeFePO, Phys. Rev. Lett. {\bf 109}, 216402 (2012).

\bibitem{Lengyel2015} E. Lengyel, M. E. Macovei, A. Jesche, C. Krellner, C. Geibel, and M. Nicklas, Avoided ferromagnetic quantum critical point in CeRuPO, Phys. Rev. B {\bf 91}, 035130 (2015).

\bibitem{Hu2015} D. Hu, X. Lu, W. Zhang, H. Luo, S. Li, P. Wang, G. Chen, F. Han, S. R. Banjara, A. Sapkota, A. Kreyssig, A. I. Goldman, Z. Yamani, C. Niedermayer, M. Skoulatos, R. Georgii, T. Keller, P. Wang, W. Yu, and P. Dai, Structural and Magnetic Phase Transitions near Optimal Superconductivity in ${\rm {BaFe_2(As_{1-\textit{x}}P_\textit{x})_2}}$, Phys. Rev. Lett. {\bf 114}, 157002 (2015).

\end{thebibliography}
\end{document}